\documentclass[fleqn,usenatbib]{mnras}

\usepackage{newtxtext,newtxmath}
\usepackage[T1]{fontenc}

\DeclareRobustCommand{\VAN}[3]{#2}
\let\VANthebibliography\thebibliography
\def\thebibliography{\DeclareRobustCommand{\VAN}[3]{##3}\VANthebibliography}

\usepackage{graphicx}	% Including figure files
\usepackage{multicol}  % Multi-column entries in tables
\usepackage{pdflscape}	% Landscape pages
\usepackage{hyperref}
\usepackage{siunitx}
\usepackage{xcolor}
\usepackage{subcaption}
\newcommand{\GG}{\mbox{$G$}}

\newcommand{\AGG}{\mbox{$A_G$}}
\newcommand{\GBP}{\mbox{$G_{\rm BP}$}}
\newcommand{\GRP}{\mbox{$G_{\rm RP}$}}

\newcommand{\RV}{\mbox{$R_{5495}$}}

\newcommand{\Teff}{\mbox{$T_{\rm eff}$}}

\newcommand{\HII}{\mbox{H\,{\sc ii}}}

\newcommand{\unum}[1]{\textcolor{black}{#1}}

\title[The ALS III Catalogue of Galactic OB stars]{The Alma catalogue of OB stars. III. A cross-match with \textit{Gaia} DR3 and an extension based on new spectral classifications.}

\author[M. Pantaleoni Gonz\'alez et al.]{
M. Pantaleoni Gonz\'alez,$^{1, 2, 3}$\thanks{E-mail: michelangelo.pantaleoni@univie.ac.at, currently at the University of Vienna.}
J. Ma{\'\i}z Apell\'aniz,$^{1}$
R. H. Barb\'a,$^{4}$
B. Cameron Reed,$^{5}$
S. R. Berlanas,$^{6,7}$
\newauthor A. Parras Rico$^{1}$
and A. Bodaghee$^{8}$
\\
$^{1}$Centro de Astrobiolog{\'\i}a. CSIC-INTA. Campus of the European Space Astronomy Centre (ESAC). E-\num[detect-all]{28692} Villanueva de la Ca\~nada. Madrid. Spain.\\
$^{2}$Department of Astrophysics. University of Vienna. \num[detect-all]{1180} Wien. Austria.\\
$^{3}$Departamento de Astrof{\'\i}sica y F{\'\i}sica de la Atm\'osfera. Universidad Complutense de Madrid. E-\num[detect-all]{28040} Madrid. Spain.\\
$^{4}$Deceased. Formerly at Departamento de Astronomía. Universidad de La Serena. \num[detect-all]{1200} La Serena. Chile.\\
$^{5}$Department of Physics (Emeritus). Alma College. Alma. \num[detect-all]{48801} Michigan. United States of America.\\
$^{6}$Instituto de Astrof\'{\i}sica de Canarias. E-\num[detect-all]{38200} La Laguna, Tenerife, Spain. \\
$^{7}$Departamento de Astrof\'{\i}sica. Universidad de La Laguna. E-\num[detect-all]{38205} La Laguna, Tenerife, Spain.\\
$^{8}$Department of Chemistry, Physics and Astronomy. Georgia College and State University. \num[detect-all]{31061} Milledgeville, United States of America.\\
}

\date{Received 2025 XXX XX. Accepted 2025 XXX XX.}

\pubyear{2025}

\begin{document}
\label{firstpage}
\pagerange{\pageref{firstpage}--\pageref{lastpage}}
\maketitle

\begin{abstract}
We present the third installment of the Alma Luminous Star (ALS) catalogue, aimed at creating the most comprehensive and clean sample of Galactic massive stars. This update extends the sample by adding approximately 2000 OB stars, incorporating astrometric and photometric data from the \textit{Gaia} Data Release 3 (DR3) alongside spectroscopic information from the Galactic O-Star Catalog based on recent ground-based spectroscopic surveys. Rigorous astrometric corrections are applied to Gaia DR3 parallaxes, proper motions, and photometry, ensuring accurate distance estimates through a Bayesian method suited to this stellar population's spatial distribution in the Milky Way. We perform some comparative analyses highlighting the improved distance accuracy over previous versions, underscore the importance of precise spectral classifications with competing catalogues, and identify areas for improvement in \textit{Gaia} DR3 effective temperature and extinction estimates for massive stars. We also address the challenges of having robust definitions for these objects. In addition, we explore the catalogue's ability to trace Galactic features such as spiral arms, spurs and OB associations (with some insights on the nature of Gould's Belt). Finally, we discuss the potential for further expanding the sample with upcoming surveys. This effort marks a significant advancement in the creation of a reliable census of Galactic massive stars, contributing to our understanding of the Milky Way's structure and star formation history. This catalogue should serve as a valuable reference for the massive star community, supporting research on stellar interiors, winds, stellar feedback, and other processes that make OB stars key to the evolution of galaxies.
\end{abstract}


%%%%%%%%%%%%%%%%%%%%%%%%%%%%%%%%%%%%%%%%%%%%%%%%%%

%%%%%%%%%%%%%%%%% BODY OF PAPER %%%%%%%%%%%%%%%%%%

\section{Introduction}

$\,\!$\indent The ``Alma\footnote{The name Alma refers to the Alma College in Michigan, where the project's originator was based, and has no relation to the Atacama Large Millimeter/submillimiter Array (ALMA) observatory.} Luminous Star'' (ALS) catalogue (\citealt{Reed03}, hereafter Paper I) was started three decades ago \citep{Reed93a} as a collection of Galactic luminous stars with available $UBV\beta$ photometric data. It was originally based on the ``Case-Hamburg'' survey of Galactic luminous stars \citet{StepSand71,Hardetal59,Stocetal60,NassStep63,Hardetal64,Hardetal65,Nassetal65} but it soon grew to include data from several hundreds of references. A parallel similar project, the Galactic O-Star Catalog (GOSC, \citealt{Maizetal04b,Sotaetal08}, see also \url{https://gosc.cab.inta-csic.es}) was also initiated by some of us two decades ago and originally concentrated only on O-type stars, but was later expanded into massive stars in general. The ALS project stayed mostly dormant for some time, until the availability of \textit{Gaia}~DR2 data allowed us to cross-match it with the ALS catalog, resulting in \citet{Pantetal21}, from now on Paper II. The original ALS data drew from a diverse range of sources, many of which provided imprecise coordinates and/or unreliable identifiers, resulting in the need to go over each source one by one to confirm the cross-match. In this modern era of automatic cross-matching done by artificial intelligence, sometimes old-fashioned human hard work is also needed. This is  especially so when the quality of the matching is poor due to diverse reasons that are sometimes unique for a specific origin: hand-written text on photographic plates with typos, scanned figures with a small unnoticed rotation included in them, IDs sketched on plates overexposed in \HII\ regions, or rounded coordinates for objects in dense clusters, to name just a few. Previous attempts at automatic cross-matching achieved an ALS-\textit{Gaia} DR2 compatibility of 30.9\% \citep{Xuetal18}, semi-automatic methods reached 63.4\% \citep{Wardetal20}, and the manual approach of Paper II resulted in 98.8\% matched (with the remaining cases being mostly stars that are too faint or too bright for \textit{Gaia} to be provided in the current data release).

In this third paper of the series we develop the ALS project along three lines:

\begin{itemize}
 \item \textbf{The availability of \textit{Gaia}~DR3 data.} Shortly after the publication of Paper II, a new (third) \textit{Gaia} Data Release became available, first in its early form as EDR3 \citep{Browetal21} and then in its final one as DR3 \citep{Valletal23}. The new data release has significantly improved astrometric \citep{Maizetal21c,Maiz22} and photometric (Weiler et al. in prep.) calibrations as well as new features such as low-resolution (BP+RP, combined as XP) spectrophotometry \citep{DeAnetal23,Weiletal23} and variability \citep{Eyeretal23,Maizetal23}. All of that allows us to obtain more accurate and precise distances to the ALS stars and to improve our differentiation between those objects that are indeed massive and those that are of intermediate or low mass. The new information allows for an improved positioning of the ALS sample in the Milky Way, thus allowing its better use for Galactic structure studies.
 \item \textbf{The inclusion of new surveys and ground-based spectroscopy.} Since the last addition of sources to the ALS two decades ago there have been several surveys that have added to our inventory of massive stars. Some of those surveys, such as the two our group is leading (Galactic O-Star Spectroscopic Survey or GOSSS, \citealt{Maizetal11}, and \textbf{Li}brary of \textbf{Li}braries of \textbf{Ma}ssive-Star High-\textbf{R}eso\textbf{l}ut\textbf{i}on Spectra or LiLiMaRlin, \citealt{Maizetal19a}) are providing spectra and spectral types for stars, some already in the ALS and some that were not. In the first case we use the new information to discriminate the nature of the existing ALS sample and in the second one to increase its sample size.
 \item \textbf{Extending the project into the future.} Given the convergence of the ALS and GOSC projects, we unify them into a single one (retaining the ALS name) and make the results available from a new web site (\url{https://als.cab.inta-csic.es}), starting with this paper. Such a unification requires some nomenclature issues that we tackle here. Also, given the expected influx of new spectroscopic data for massive stars from the above cited projects and others such as WEAVE \citep{Jinetal24} and 4MOST \citep{Chiaetal19,Cionetal19}, the ALS sample should increase significantly in the near future, not only in the Galaxy but also in the Large Magellanic Cloud (LMC), Small Magellanic Cloud (SMC), and Magellanic Bridge (MB). In this paper we discuss how we will handle this, including the extension of the ALS from a single-galaxy to a three-galaxy project.
\end{itemize}

This paper has been written simultaneously with three others, from which it feeds in part. The first one is the last entry of the GOSSS project (GOSSS IV, Ma\'{\i}z Apell\'aniz et al. in prep.), which extends the sample of Galactic O stars to over 1000. The second is a new spectral classification grid for B stars \citep{Neguetal24}. The third one is another new spectral classification grid (Ma\'{\i}z Apell\'aniz et al. in prep.), in this case for O and early-B stars, that feeds from GOSSS and LiLiMaRlin but also from projects in the Magellanic Clouds such as the VLT-FLAMES Tarantula Survey (VFTS, \citealt{Evanetal11a,Walbetal14}) and XShootU \citep{Vinketal23}.

In the next section we describe how we have built the third version of the ALS catalog, describe its nomenclature, and briefly explore its contents. After that, we present some results on distances, a comparison with \textit{Gaia}-based effective (\Teff) and extinction measurements, and the Galactic structure in the solar neighbourhood. We end with a description of the future lines of work of the ALS project. 

%%%%%%%%%%%%%%%%%%%%%%%%%%%%%%%%%%%%%%%%%%%%%%%%%%
\section{Data and methods}
\label{subsec:buildup}

\subsection{The importance of defining OB stars and the need for a purely Galactic OB star catalogue}
\label{subsec:needOB}

Due to \textit{Gaia}'s outstanding astrometric capabilities, the field of Galactic cartography is currently undergoing a revolution. As young massive stars are recognised as an important tracer of the star-forming regions and overall spiral arm structure of the Milky Way (\citealt{Morgetal53b}), it is of the utmost importance that we obtain accurate classifications and operate with strict definitions. An OB star is classically defined as a massive star ($M \gtrsim 8 M_{\odot}$) of either O or B spectral type. This approximately translates to stars of spectral subtype B2 or earlier for dwarfs, B5 or earlier for giants, and B9 or earlier for the supergiants \citep{Morg51,Maizetal24c}. It is an approximation because there are exceptions; for example, postAGB stars are low-mass objects that briefly appear to be OB giants or supergiants, hot subdwarfs can have spectral characteristics similar to OB stars near the main sequence and, of course, binary evolution can produce exceptions to almost anything that can be said about single-star evolution, sometimes giving low-mass systems OB characteristics (colors match and luminosity can be deceptively large if the system is not resolved). However, those contaminants are a small fraction of the total and, in some cases, can be distinguished from real massive stars if the available spectroscopic data is analysed with care.

Due to the shape of the IMF and their short lifespans, OB stars (as defined here) are actually a minority in the set of Galactic O- and B-type stars, which in general are not massive enough for this definition to apply. Most of them are actually B2.5-B9 dwarfs, which are of intermediate mass (\citealt{Martetal05a}), a fact ignored in several recent papers (see Table~\ref{tab:oba_catalogs} and the following discussion for details). This distinction is scientifically relevant because OB stars can ionize the surrounding ISM (contributing to feedback); carve large shells with their winds, eruptions, and explosions (\citealt{Fieretal16,Vink22}); trace star-forming regions (\citealt{Morgetal53b}), play a key role in the Galaxy’s chemical evolution (\citealt{Nomoetal13}); and generate runaway and walkaway stars through dynamical interactions (\citealt{Poveetal67}) and supernova explosions (\citealt{Blaa61}). In contrast, non-massive O- and B-type stars generally lack these characteristics. Relaxing these conditions frequently results in ``OB'' samples dominated by false positives. Additionally, false negatives are common when classifying OB stars with photometry, as interstellar extinction and reddening often cause OB stars to be confused with later types. This issue is exacerbated by their higher luminosity, which allows them to be detected at greater distances, where extinction tends to be more significant.

As the demand for OB star identifications grows, many new and large OB(A) star catalogs are being created using \textit{Gaia} DR3 spectrophotometry or ground-based spectroscopic surveys. However, this is sometimes achieved not by adding more data or improving the selection criteria but, on the contrary, by relaxing aspects of the classical definition, which allow non-massive stars to enter the sample. It is therefore essential to highlight the strengths and limitations of the ALS III catalogue in this context and to critically asses the evolving landscape of new hot-star samples.

\begin{table}
    \centering
    \begin{tabular}{l r l}
        \hline
        Catalogue reference & Size & Definition used \\
        \hline
        \cite{Poggetal21} & \num{606219} & Upper Main Sequence stars\\
        \cite{Zarietal21} & \num{435273} & OBA stars \\
        \cite{Khaletal24} & ${\sim}$\num{375 000} & B-type candidates \\
        \cite{Xianetal22} & \num{332172} & OBA stars with $T_{\text{eff}} \gtrsim 7.5$ kK\\
        \cite{LiBinn22} & \num{46916} &  ``OB stars'' with $T_{\text{eff}} \gtrsim 10$ kK\\
        \cite{Liuetal24} & \num{27643} & ``OB stars''\\
        \cite{Quinetal25} & \num{24706} & B9.5 or earlier-type stars\\
        \cite{Liuetal19b} & \num{22901} & ``Normal OB stars''\\
        ALS III (this work) & \unum{\num{15542}} & OB stars (non-binary GLS)\\
        \cite{Chenetal19b} & \num{14880} & OB stars down to B3V\\
        \cite{Xuetal21} & \num{9750} & O-B2 stars from the literature\\
        \hline
    \end{tabular}
    \caption{Summary of recent catalogs with OB stars.}
    \label{tab:oba_catalogs}
\end{table}

In Table~\ref{tab:oba_catalogs} we summarise the definitions employed in some of the most recent hot-star catalogues in the literature. In general there are three main tendencies;

\begin{itemize}
    \item Catalogues like the ones from \citet{Poggetal21}, \citet{Zarietal21}, \citet{Xianetal22} and \cite{Khaletal24}, aim for higher completeness and thus adopt definitions that encompass hot non-OB stars as well, which makes them consistently larger by at least one order of magnitude compared to the ALS III. For this comparison we have selected only the stars in the catalogue with GLS prefixes that are not inherited from nearby GLS-prefixed companions (stars following the G2, G3, G4 and G5 selection criteria shown in Fig.~\ref{fig:flowchart}, which is explained in detail in section \ref{subsec:prefix}). The Upper Main Sequence (``UMS'') sample in \citet{Poggetal21} and the OBA catalogue from \citet{Zarietal21} were built with broad selection criteria, as a first attempt to explore Galactic cartography with potentially young stars in \textit{Gaia}. Both were extremely successful at that and as a result large scale structures, like spiral arms, are now clearly visible in their smoothed density maps. But using these catalogues to gain detailed knowledge of the young massive population or to explore individual stellar groups can be problematic, since both the UMS sample (which aimed for stars more luminous than B3V) and the OBA star sample (which aimed for stars of earlier types than B7) are effectively dominated by late subtype B stars and A-type stars. For the purpose of studying massive stars these are A-type star catalogues contaminated by a few OB stars. In \cite{Khaletal24} the sample is cleaner (and smaller) but the definitions used (without considering the methods) also allow for late B-type stars. The recent catalogue by \cite{Quinetal25} defines their OB sample as stars of spectral type B9.5 or earlier (i.e., with $T_{\text{eff}} \gtrsim 10$ kK). However, their careful analysis, based on fitting the spectral energy distributions (SEDs) of stars from multiple photometric surveys, reduces the number of candidates to a size comparable to that in this work. Therefore most of these stars, though young and thus relevant for mapping purposes, are not massive. In \citet{Xianetal22} the concept of OBA stars is relaxed to the point of allowing stars with temperatures of $7.5$ kK (way below the ${\sim}20$ kK expected for the cooler main sequence massive stars). These notes are directed to the general reader, which might be tempted, by the large numbers involved in these catalogues, to extrapolate strong statistics for young massive stars, but it is not a criticism of these studies, which in fact are honest with their results and completely coherent with the definitions employed.
    \item Catalogues like \citet{Chenetal19b} (which allows for B3V stars to be included in the definition of OB stars), \citet{Xuetal21} and the one presented in this work, attempt to avoid contamination from later types and distill their samples by gathering spectroscopic identifications from generally trusted sources in the literature. Both \citet{Chenetal19b} and \citet{Xuetal21} rely on the immense compilation of spectral classifications of the literature by \citet{Skif14} but also the original ALS and GOSC spectral classifications. The weakness of these catalogues (including the ALS III) is that their size tends to be smaller. But again, they should be smaller since OB stars are extremely rare and having low-number statistics is preferable to having good statistics about the wrong stellar population. Of these category of catalogues, the ALS III aims to be the largest and more consistent, currently available.
    \item Catalogues like \citet{Liuetal19b, Liuetal24} and \citet{LiBinn22} are middle sized catalogues of purportedly ``OB stars'', but the definitions and methods used in these papers incorrectly imply that ``OB stars'' are simply O-type and B-type stars. In the case of \citet{LiBinn22}, even temperatures as low as $10$ kK are considered permissible, and these are taken from the StarHorse general catalogue (\citealt{Andeetal22}), which explicitly warned the reader by stating that ``\textit{our results for very massive stars are very likely to be unreliable in most cases}''. In \citet{Liuetal24}, \num{27647} stars are said to be ``Normal OB stars'', selected with the help of LAMOST DR7 low-resolution spectra. Of these, \num{16110} seem to have good quality evaluations, which were obtained with the MKCLASS automatic spectral classifier (\citealt{GrayCorb14}). In figure 7 of their paper it is clear that more than ${\sim}80\%$ of these are not OB stars in the usual sense.
\end{itemize}

With this, we want to make the reader aware of the benefits and problems associated with different approaches when defining OB stars, and the confusion that might ensue from the inconsistent use of the ``OB'' label across the recent literature. In Section \ref{subsec:maps}, we discuss how adopting different definitions for what OB stars are, has a significant impact on Galactic cartography.

\subsection{Building the new version of the ALS catalogue}
$\,\!$\indent The final goal of the ALS catalogue is to include entries for all known massive stars and their companions, excluding in principle red supergiants (which may be included at a later stage). This goal requires large-scale spectroscopic surveys that are currently not available but are expected to be so in the next decade. The immediate goal of this paper is to continue with this process, which we started in Paper II, as a stepping stone for future papers of the series. This progression necessitates careful deliberation when considering the inclusion of a new object or the exclusion of a star in ALS III (the version presented in this paper) that was previously cataloged in ALS II (the version associated with Paper II). The reason is that we want to minimize future nomenclature changes, which will be inevitable to some degree nonetheless, while adding as many new stars as possible. In this subsection we list the ways in which we have added new stars to the sample and in the rest of this section we describe the nomenclature, the data reduction for \textit{Gaia} astrometry and photometry, and the catalogue contents.

For the construction of the ALS III catalogue, new stars have been added to the ALS II catalogue through different means (see upper part of the flow chart in Fig.~\ref{fig:flowchart});

\begin{itemize}
 \item \textbf{ALS (P)}: A programmatic error during the upload of the original ALS tables to the CDS services resulted in the catalogue being truncated, leaving only the first \num{18693} entries publicly accessible. This issue went unnoticed even during the preparation of the ALS II, 18 years later. In this paper, we finally include the \unum{\num{1119}} previously missing entries from this inadvertently private portion of the ALS (hereafter referred to as ALS (P)). To preserve continuity, we refrain from removing any entries or altering identifiers, as some of these stars were cited in subsequent publications following Paper I. As in Paper II, we crossmatched almost every source with \textit{Gaia}, examining the available references and searching for astrophotometrically similar sources within a few arcseconds and small magnitude differences. This careful analysis allowed to identify \unum{\num{10}} additional duplicates in the original ALS and expanded the reaches of the catalogue to highly obscured clusters like Arches, Quintuplet, Galactic Center, and W31.
 
 \item \textbf{GOSC:} As previously stated, the ALS III is now encompassing the GOSC project, which is fed by several spectroscopic surveys, like GOSSS and LiLiMaRlin (which is itself fed by spectroscopic surveys like OWN, IACOB, NoMaDS and CAFÉ-BEANS). In particular GOSC contributes with \unum{\num{3468}} verified spectral classifications, both published and unpublished (which will be available in the new ALS web site). The majority of those spectral classifications come from the first three major GOSSS papers (I: \citealt{Sotaetal11a}, II: \citealt{Sotaetal14}, and III: \citealt{Maizetal16}) but others come from subsequent papers, most notoriously the dedicated studies of Cygnus~OB2 (\citealt{Berletal20}) and Carina~OB1 \citep{Berletal23}. However, most of them are still unpublished at this stage, something that will be partially remedied with the upcoming GOSSS IV (Ma\'{\i}z Apell\'aniz et al. in prep.) and a new paper of the Villafranca series on Carina~OB1 (Molina Lera et al. in prep.). Overall, GOSC has increased the number of entries in the ALS by \unum{\num{506}}.
  
 \item \textbf{GWRC:} The Galactic Wolf Rayet Catalogue (here GWRC; \citealt{RossCrow15}) in its most recent version (\unum{v1.30}), Contains \unum{\num{679}} confirmed Wolf Rayets. We adopted \unum{\num{678}} of them, and reject what we believe to be an internal duplicate (WR 102-4 is the same as WR 102ae). Some of these Wolf Rayets were present in the original ALS and ALS (P), but \unum{\num{490}} cases have expanded the list for the ALS III.
\end{itemize}

The planning for future additions is described in the last section of this paper. In summary, the idea is to use \textit{Gaia} astro-photometric information to yield high-quality candidates for which subsequent spectroscopy will provide verified spectral classifications. Such a scheme is the natural evolution of the procedure in this paper but with a much larger sample, which we expect to grow by an order of magnitude in the following decade.

\subsection{Nomenclature}
The names in the ALS III are composed of a prefix (``GLS'' or ``ALS''), a number (which reflects the complex history of catalogue expansions) and a string suffix (that, when present, allow us to discern between components in multiple systems).

\subsubsection{ALS prefixes}
\label{subsec:prefix}
\begin{figure*}
    \centering
    \includegraphics[width=0.88\textwidth,keepaspectratio]{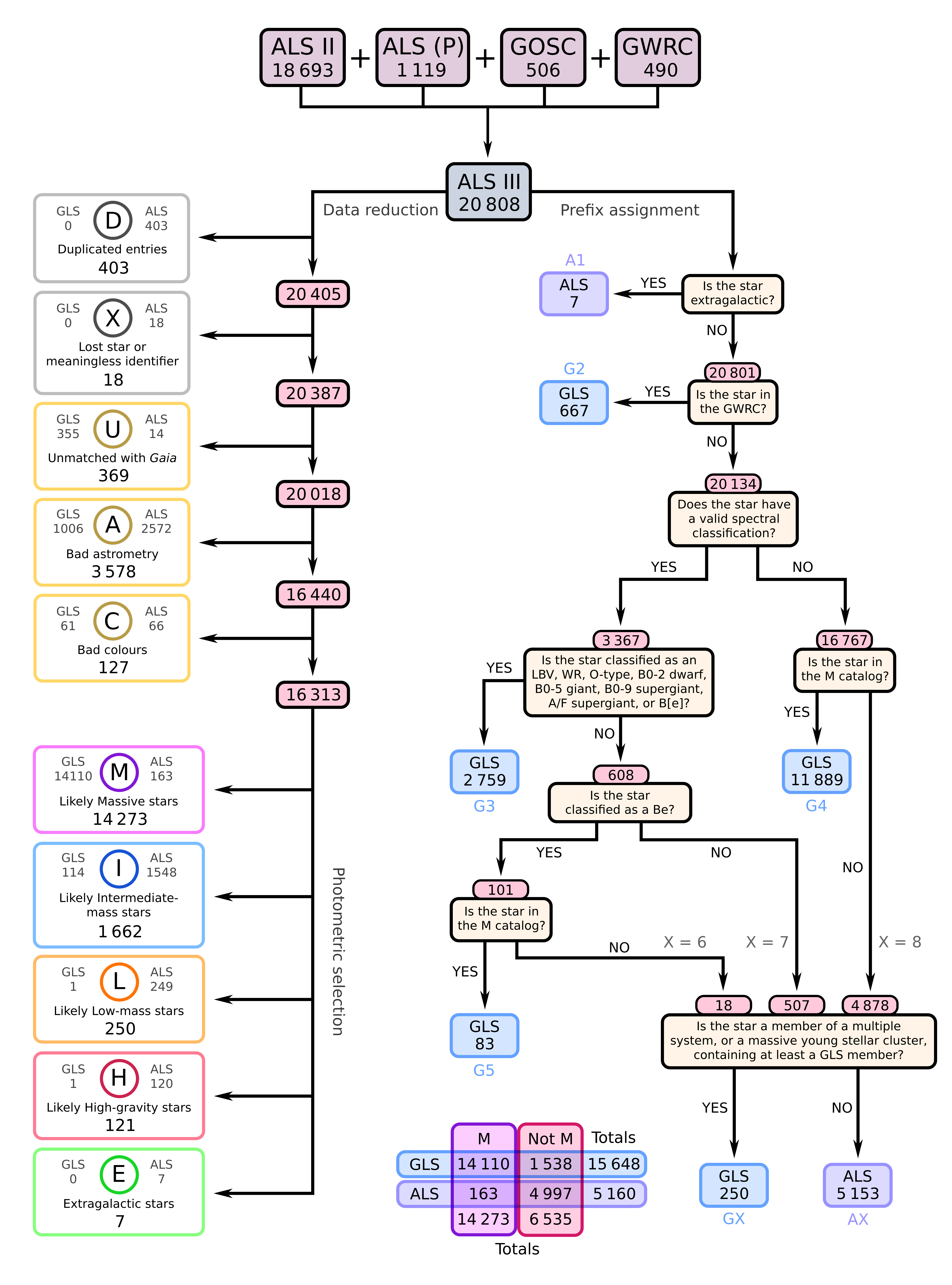}
    \caption{Complete flowchart illustrating (upper part) the catalogue construction process leading to the ALS III (described in section \ref{subsec:buildup}), (left branch) the ALS III data reduction procedure and internal catalogue partition (described in sections \ref{subsec:datared} and \ref{subsec:cats}, respectively), and (right branch) the GLS/ALS prefix assignment decision tree (described in section \ref{subsec:prefix} and table \ref{tab:prefixsummary}). A confusion matrix comparing the GLS/ALS prefixes (which in general convey information about validated spectroscopic classifications) against the M internal catalogue (which conveys information about the photometric detection of likely massive stars) can be found at the bottom. The proportions of ``GLS'' and ``ALS'' prefixes within each internal catalogue (from both data reduction and photometric selection) are displayed on the upper corners of each box.}
    \label{fig:flowchart}
\end{figure*}

With the catalogue integrating data from a wide variety of sources spanning nearly a century of developments in the field, it is essential to establish clear criteria for reliably identifying true Galactic luminous stars and at the same time, preserving entries from the original ALS, which occasionally (albeit rarely) includes lower-mass stars (as shown in Paper II and in this work). The identifier prefix in the ALS catalogue, historically limited to ``ALS'' (for Alma Luminous Star), has now been expanded to include the ``GLS'' (Galactic Luminous Star) prefix. This new label reflects a higher degree of confidence in classifying any star as a massive OB star; meaning a progenitor with a mass exceeding $8 M_{\odot}$ (in section \ref{subsec:needOB} we discussed the importance of having strict definitions in regards to OB stars). Distinguishing such cases through an updated naming convention requires applying both photometric and spectroscopic criteria, which are shown in the decision tree of Fig.~\ref{fig:flowchart}.

In essence, we want to give the ``GLS'' prefix to confirmed Wolf Rayet stars (G2 path in the decision tree) and to stars that have reliable validated spectroscopic identifications as OB stars; meaning any O-type star, subtypes B0 to B2 for dwarfs, B0 to B5 for the giants, B0 to B9, plus A- and F-type stars for the supergiants and any B[e] star (through the G3 path of the decision tree). But, given the current scarcity of accurate spectral classifications for the vast majority of the ALS III, other ways of assessing the status as a massive star are needed. For this, we follow the selection criterion used to define what we call the ``M sample''; the subset of stars with good‑quality \textit{Gaia} DR3 astrometry and photometry that lie above the extinction track for a main sequence $T_{\text{eff}} = 20$ kK star (see \ref{subsec:cats} for full details). This M criterion is used as a secondary confirmation for stars that have validated Be classifications (G5 path) or any star that simply lacks any validated spectral classification (G4 path). Besides this, $7$ extragalactic stars are carried from the original ALS catalogue and we avoid giving them a ``GLS'' prefix (since they are non-Galactic) even if they may comply with the criteria (these follow the A1 path and mantain their ``ALS'' prefix). In a future paper we will generate codes for stars in the LMC, SMC, and MB.

The remaining \unum{\num{5403}} stars originate from three possible paths: those confirmed as Be stars through validated spectra but not identified photometrically (path 6), those with validated spectra confirming them as non-massive stars, regardless of whether they belong to the M internal catalogue (path 7), and those without validated spectra and deemed photometrically inapt (path 8). While stars following path 7 are definitively non-massive, paths 6 and 8 leave open the possibility of promotion to ``GLS'' status in the future, as improvements in photometry and astrometry, particularly from upcoming \textit{Gaia} data releases, may enable more stars to transition into the M internal catalogue (particularly from the A, C, and I internal catalogues, as described in sections \ref{subsec:datared} and \ref{subsec:cats}).

A final criterion is applied when assigning prefixes to stars following paths 6, 7 and 8. We grant the ``GLS'' status to \unum{\num{144}} stars that are members of Westerlund 1, Westerlund 2, NGC 3603, the Quintuplet cluster, the Arches cluster and the Galactic Center cluster. All bright stars in these clusters are considered to be massive stars, as no low-mass star is optically accessible at such extinctions. Additionally, the ``GLS'' prefix is assigned to \unum{\num{106}} stars that are companions of stars already designated ``GLS'' based on their own merits. This is purely done to help harmonize names and allow better readability when looking at members of multiple systems. Consequently, these remaining prefixes are distributed among stars that follow the 6, 7 or 8 paths ending in either ``GLS'' or ``ALS''.

This results in a total of 11 distinct paths for prefix assignment (A1, G2, G3, G4, G5, A6, G6, A7, G7, A8, and G8), which are documented in the tables under the ``Pflag'' column (from \textit{prefix}). The bulk of the sample is dominated by the G4 (Galactic massive stars confirmed with photometry, \unum{\num{11889}} objects), A8 (non-M stars without an alternative confirmation, \unum{\num{4636}} objects), and G3 (spectroscopically confirmed Galactic massive stars, \unum{\num{2759}} objects) paths.  

Ideally, all stars with a ``GLS'' prefix would be Galactic massive stars (and those with ALS prefix something else). In reality, we expect to have both false positives (non-massive stars that have received a GLS prefix) and false negatives (massive stars with an ALS prefix). False positives will be dominated by G4-path objects, for which we have no spectral classification, that are close to the 20~kK extinction track but are intermediate-B dwarfs. This can happen because e.g. a somewhat evolved B4~V star can be brighter than a B2~V in the zero-age main sequence (ZAMS) or because of the different extinction tracks that can be followed as a function of \RV\ (see \citealt{Berletal23} for examples). Another source of false positives will be objects with anomalous evolutionary tracks such as fast rotators, as for them their evolved effective temperatures may be higher (and their spectral types earlier) than expected for their mass and age. False negatives will be dominated by A8-path stars with bad colours or astrometry (subcatalogs C and A). Future ALS editions will change some of the prefixes through two mechanisms: by obtaining new spectral classifications  and by the improvements introduced by future \textit{Gaia} data releases.

\begin{table*}
    \centering
    \begin{tabular}{l c l}
        \hline
        Pflag & Number & Description \\
        \hline
        A1 & \unum{\num{7}} & ALS-prefixed as these are the extragalactic (LMC+SMC) stars present in the original ALS catalogue\\
        A6 & \unum{\num{17}} & ALS-prefixed as the star is a Be below the extinction track for a $T_{\text{eff}} = 20$ kK MS star\\
        A7 & \unum{\num{500}} & ALS-prefixed as it is not a Be nor an OB star (LBV, WR, O-type, B0-2 dwarf, B0-5 giant, B0-9 or A/F supergiants, or B[e])\\
        A8 & \unum{\num{4636}} & ALS-prefixed as the star has no classification and it is below the extinction track for a $T_{\text{eff}} = 20$ kK MS star (includes D and X)\\
        G2 & \unum{\num{667}} &  GLS-prefixed via Wolf-Rayet Catalogue\\
        G3 & \unum{\num{2759}} &  GLS-prefixed via spectral classification\\
        G4 & \unum{\num{11889}} & GLS-prefixed via the color-absolute magnitude diagram (CAMD; M catalogue)\\
        G5 & \unum{\num{83}} & GLS-prefixed via Be star classification and CAMD (M catalogue)\\
        G6 & \unum{\num{1}} & GLS-prefixed via binary criterion and the star is a Be\\
        G7 & \unum{\num{7}} & GLS-prefixed via binary criterion and the star is not a Be\\
        G8 & \unum{\num{242}} & GLS-prefixed via binary/cluster criterion and the star has no spectral classification\\
        \hline
    \end{tabular}
    \caption{Summary of the paths and selection criteria followed by the prefix assignment process for the ALS III catalogue (see Figure \ref{fig:flowchart})}
    \label{tab:prefixsummary}
\end{table*}

\subsubsection{ALS numbers}
The ALS number identifiers only contain information about the long history of the project, and are never modified between catalogue releases. The original ALS catalogue has numbers that run from star \num{1} to star \num{20171}, with two large gaps where no number is assigned. In particular, the range from \num{1} to \num{5132} was assigned for the ``Case-Hamburg'' southern stars, from \num{6001} to \num{13390} to the ``Case-Hamburg'' northern stars, and from \num{14001} upward (up to \num{20171} considering the available original ALS, and continuing without additional gaps up to the number \num{21290} considering the private part of the catalogue that we are finally uncovering) for supplemental OB stars collected from nearly a thousand references of the literature. We have now assigned new ALS numbers, from \num{22000} to \unum{\num{22469}} for additional stars gathered from GOSC, and identifiers from \num{23000} to \unum{\num{23490}} from the GWRC expansion. This means that, in general, ALS numbers below \num{22000} refer to entries of the original ALS (the public part and also ALS (P)), numbers of the form \num{22} XXX are for new stars, added from GOSC, and numbers of the form \num{23} XXX are for new stars  added from the GWRC.

The exceptions to these numbers are previously unrecognized multiple systems that GOSC identifies as having OB companions. In this case, newly added stars from GOSC can have ALS numbers from the set of identifiers of the original ALS catalogue (including ALS (P)), but will have different letter suffixes for different components. This implies that an ALS number is not enough to uniquely identify a star in the catalogue (but the number plus the component suffix is).

\subsubsection{ALS suffixes}
We add an ALS suffix of the type A, B, Aa, Ab, and so on following the classical nomenclature for the hierarchical architecture of multiple systems. In most cases we follow the component nomenclature of the Washington Double Star (WDS) catalogue \citep{Masoetal01}, but there are also special cases.

Since we ought to obey the rule that no previously published ALS number is to be changed between releases, we are forced to make an exception to the suffix system also, which implies that not all members of a multiple system will necessarily have a suffix. For example, binary systems where both components had previously assigned numbers (those numbers been different) remain with those numbers and are not assigned any suffixes, even if they are components of the system. For example, for the 1 Cam system, the ALS numbers \num{7984} and \num{7982} were assigned to the A and B components respectively (no suffixes are given to these components because they are uniquely identified with different numbers since paper I), but in systems like HD 203 938, where one component was in the original ALS but the other is been included here for the first time, we can give the same ALS number to both stars and add a suffix to each one of them (11 839 A and 11 839 B for this particular case).

A consequence of applying this exception is that not all newly added GOSC stars have ALS numbers of the form \num{22} XXX, since some are companions of stars in the original ALS and thus are assigned the old number (in the ranges covered by the original ALS and the ALS (P)) plus a suffix (to differentiate it from the other components).

\subsubsection{GOSC and ALS nomenclatures}

There are two approaches to survey nomenclature. Surveys with small samples ($\lesssim 10^6$ objects) such as HD or NGC frequently use a number that forms a sequence from 1 to the total number of objects. Surveys with larger samples, on the other hand, use a longer number (or combination of) that may encode coordinates (e.g. 2MASS) or simply be a running number with possible gaps (e.g. \textit{Gaia}). The first type are easier to memorize, which leads to the second type commonly referred to as ``telephone numbers''. For example, it is reasonable to remember that Sirius is HD~\num{48915} but preferring 2MASS~J06450887$-$1642566 as a mnemonic would not be serious. The second type, on the other hand, has the advantage of being more thorough and for allowing additions and subtractions to the sample more easily.

In this paper we are merging the ALS and GOSC projects, one having a nomenclature system of the first type and the other one of the second, as GOSC uses a prefix of the type GOS/GBS/GAS/GWS (Galactic O/B/A/WR Star) followed by the Galactic longitude and latitudes written as LLL.ll$\pm$BB.bb (with the values rounded to the nearest cell center) and an index number $\_$XX starting with 01 common to stars of all spectral types (e.g. if a GOS~111.11$+$11.11$\_$01 star exists and a B star in the same coordinate cell is added to the catalog, that receives the ID GBS~111.11$+$11.11$\_$02). We plan to maintain the dual nomenclature for practical reasons. First, ALS numbers are easier to memorize and can provide a useful permanent ID to objects that currently have only ``telephone numbers'' or worse (e.g. IDs like NGC XXXX YYY for an object in a stellar cluster with several different surveys and conflicting YYY numbers, leading to confusion). Second, because the GOSC nomenclature assigns a fixed value (excluding the prefix) to each object, while the ALS one changes as components are included. Third, because the GOSC nomenclature allows for an immediate ID to be assigned to an object, which may not be published in months or years. Indeed, as described below, the ALS web service includes two different pages, one for membership and the other for spectral types, which depend on a common database. That database has two copies, a development (private) and a production (public), with the first one constantly changing but with the changes transferred to the second one only when the data are published in a paper (such as this one). Fourth, because the two samples have differences (e.g. a late B dwarf may receive a GBS ID but not an ALS one). Fifth, because for historical reasons both IDs are useful.  

One (near) future modification to the ALS and GOSC nomenclatures will be the incorporation of LMC, SMC, and MB stars, something that has already happened at the development database level. The idea there is to maintain the numbering but to change the prefixes to LLS and LOS/LBS/LAS/LWS for the LMC, to SLS and SOS/SBS/SAS/SWS for the SMC, and to BLS and BOS/BBS/BAS/BWS for the MB.

\subsection{Data calibration and catalogue contents}
\subsubsection{Correcting parallaxes}
\label{subsec:parallax}
Astrometry in \textit{Gaia} DR3 remained unchanged with respect to \textit{Gaia} EDR3 \citep{Valletal23}. We corrected the parallax values $\varpi$ (retrieved from ESA's Gaia Archive) by the zero-point systematics $\varpi_0$, described in \citet{Maiz22}, which improved the parallax bias recipe in \citet{Lindetal21b} for bright stars, following the equation

\begin{equation}
    \varpi_c = \varpi-\varpi_0
\end{equation}

Moreover, \textit{Gaia} DR3 significantly underestimates the catalogue uncertainties in parallax, $\sigma_{\text{int}}$. To correct these internal random uncertainties we scale them up by a constant $k$, whose value is calculated following \citet{Maiz22}. Then, an unaccounted systematic uncertainty of $\sigma_{\text{sys}} = 10.3\, \mu\text{as}$ \citep{Maizetal21c} is added in quadrature according to Eqn. 1 in \citet{Fabretal21a} to yield an external uncertainty of

\begin{equation}
    \sigma_{\text{ext}} = \sqrt{(k\sigma_{\text{int}})^2+\sigma_{\text{sys}}^2}
    \label{eq:extunc}
\end{equation}

The resulting $\varpi_c \pm \sigma_{\text{ext}}$ values improve the accuracy of \textit{Gaia} DR3 $\varpi \pm \sigma_{\text{int}}$ values for single sources by a significant amount. Papers that directly make use of the \textit{Gaia} DR3 values are prone to make overconfident assessments of parallax uncertainties and systematically overestimate distances for bright stars.

Note that the corrected parallaxes and distance estimates presented in the ALS III are derived for single \textit{Gaia} sources and can therefore be improved in cases where information about membership to a stellar group is considered. An important advice for those interested in getting parallaxes for groups of closely separated sources is to avoid calculating any simple uncorrelated average, and instead follow a recipe (like the one found in \citealt{Maizetal21c}), that adequately accounts for the strong angular covariance between closely separated sources. The reader is referred to the papers of the Villafranca project (\citealt{Maizetal20b,Maizetal22b}; Molina Lera et al. in prep.) for more precise \textit{Gaia} distances to a number of Galactic OB groups, with an updated listing given in \citet{Maizetal24b}. Once the Villafranca project increases it sample, we plan to incorporate its distances into future versions of the ALS project.

\subsubsection{Correcting proper motions}
\label{subsec:pm}
Corrections to \textit{Gaia} DR3 proper motions are implemented for bright stars, following the recipe of \citet{CanGBran21}.

Since proper motion uncertainties are also underestimated in \textit{Gaia} DR3, we apply equation \ref{eq:extunc} to them. Note that this equation was used for parallaxes but can also be used in this context as long as the $k$ values are correctly derived from \citet{Maiz22} and the systematic uncertainty used is the $\sigma_{\text{sys}} = 23\,\mu\text{as}/\text{a}$ estimated in \citet{Lindetal21a}.

We also transformed these proper motions from ICRS to Galactic coordinates by following \citet{Hobbetal22}, and by using the full $10\!\times\!10$ astrometric covariance matrix that can be built from the correlations between $\alpha_{\text{ICRS}}$, $\delta_{\text{ICRS}}$, $\varpi$, $\mu_{\alpha}$ and $\mu_{\delta}$, present in the \textit{Gaia} DR3 catalog. To derive the corrected uncertainties for each component of the proper motion in Galactic coordinates ($\sigma_{\mu_{l}}$ and $\sigma_{\mu_{b}}$) we first scaled the uncertainties by the $k$ factor, then we applied the coordinate transformations and finally we add the systematic uncertainty in quadrature (basically breaking equation \ref{eq:extunc} in two steps and performing the coordinate transformation as an intermediate step).

\subsubsection{Correcting the photometry}
\label{subsec:calibphot}
Epoch-averaged photometry in \textit{Gaia} DR3 remained unchanged with respect to \textit{Gaia} EDR3, with the exception of those cases were \citet{Rieletal21} correction for the $G$ band applied. Since that correction was implemented in \textit{Gaia} DR3, we don't need to apply it anymore, but other corrections are still necessary to archive an even greater photometric precission.

In \citet{MaizWeil18} a full recalibration of \textit{Gaia} DR2 $G$+$G_{\text{BP}}$+$G_{\text{RP}}$ photometry was implemented, using reliable standards in HST spectrophotometry, by following the techniques developed in \citet{Weil18}. The photometric corrections at that time were used in the construction of the ALS II. A similar recalibration has been recently worked out for \textit{Gaia} EDR3 photometry, by the same collaboration (\citealt{MaizWeil24}; Weiler et al. in prep.). These corrections are applied to the $G$ band for all stars with a \textit{Gaia} DR3 counterpart in ALS III, resulting in the corrected magnitude, $G'$.

To calculate the uncertainties in \textit{Gaia} photometry, we follow the CDS recipe (using the mean flux and its corresponding uncertainty in the $G$, $G_{\text{BP}}$ and $G_{\text{RP}}$ bands), applying the following formulae

\begin{equation}
    \sigma_{\text{pb}} =\sqrt{\left [\frac{-2.5}{\ln 10}\left (\frac{\text{\small phot\_pb\_mean\_flux\_error}}{\text{\small phot\_pb\_mean\_flux}}\right )\right ]^2+\sigma_{\text{pb}_0}^2}
\end{equation}

where pb = G, BP and RP is the index representing each ``photometric band'', while $\sigma_{\text{pb}_0}$ represents the minimum photometric uncertainty that needs to be considered, following the values reported in \citet{MaizWeil24}:

\begin{equation}
\begin{aligned}
\sigma_{\text{G}_0} =&
      \begin{cases}
      4.8 \;\text{mmag}\;  \Longleftrightarrow \;  G \leq 13.00 \;\text{mag}\\
      3.9 \;\text{mmag}\;  \Longleftrightarrow \;  G > 13.00 \;\text{mag}\\
      \end{cases}\\
\sigma_{\text{BP}_0} =&
      \begin{cases}
      3.8 \;\text{mmag}\;  \Longleftrightarrow \;  G \leq 10.87 \;\text{mag}\\
      4.0 \;\text{mmag}\;  \Longleftrightarrow \;  G > 10.87 \;\text{mag}\\
      \end{cases}\\
\sigma_{\text{RP}_0} =&
      \begin{cases}
      5.7 \;\text{mmag}\;  \Longleftrightarrow \;  G \leq 10.87 \;\text{mag}\\
      3.4 \;\text{mmag}\;  \Longleftrightarrow \;  G > 10.87 \;\text{mag}\\
      \end{cases}
\end{aligned}
\end{equation}

We assume the uncertainty in $G$ to be the same as in the corrected version, $G'$.

\subsubsection{Calculating distances}
\label{subsec:dist}
It is well known that the inverse of the measured parallax is not an unbiased estimator for the distance to a star, due in part to the non-linearity of the transformation and the asymmetry of the resulting probability distribution (\citealt{LutzKelk73, Smit03, Lurietal18}). To mitigate these biases we can make use of Bayesian inference, combining the observed parallax with a prior distance distribution, allowing us to update our \textit{``degrees of belief''} about the plausible distances. Currently, the most cited work doing this for \textit{Gaia} (E)DR3 data is \citet{Bailetal21}, which uses a prior distribution based on the GeDR3 mock catalogue\footnote{The name GeDR3 should not be confused with Gaia EDR3. The former refers to simulated data that was generated to mimic the expected contents of the early data release (the latter) prior to its actual publication.} from \citet{Rybietal20}. However, there are two primary reasons why these distance estimates had to be improved for the ALS III:

\begin{enumerate}
    \item Although the parallax zero-point corrections from \citet{Lindetal21b} were applied in \citet{Bailetal21}, they did not incorporate the more recent improvements for bright stars outlined by \citet{Maiz22}, which are relevant to our sample. Furthermore, the now well-documented underestimation of the total parallax uncertainty (including both random and systematic components) in \textit{Gaia} (E)DR3 (\citealt{Maizetal21c}) was not considered in that work either, as this analysis was unavailable at the time of publication.
    \item The chosen prior distance distribution in \citet{Bailetal21} is highly detailed, having both a longitudinal and latitudinal dependence. But the prior is informed by the general population observed by \textit{Gaia}, which is overwhelmingly over-represented by late-type dwarfs (in the near field) and red giants (at larger distances). Since our focus is on young, massive stars, the prior must reflect the distinct spatial distribution of OB stars, including the shorter scale height of the Galactic thin disc. As the authors of that paper note: ``\textit{If one focuses on a restricted set of stars with some approximately known properties, it will be possible to construct more specific priors and to use these to infer more precise and more accurate distances}''.
\end{enumerate}

To improve the distance estimates we address both issues by following the same steps as in Paper II, using a prior based on the spatial distribution of OB stars derived from \textit{Hipparcos} data \citep{vanL07a}, modeled as an infinite, thin, isothermal, self-gravitating disc with a scale height of $h_d = 31.8$ pc, along with an additional Gaussian halo component (primarily consisting of massive runaway stars) with a scale height of $h_h = 490$ pc, as viewed from the Sun, which is assumed to be located at $z = 20$ pc above the mid-Galactic plane \citep{Maiz01a, Maiz05c, Maizetal08a}. For objects at $|b| > 1^{\circ}$ away from the Galactic plane, we use a standard mixture between the disc and halo components of the model, with the halo contributing a $3.9\%$. For stars within $1^{\circ}$ of the Galactic plane, and to prevent the improper behavior of the prior at low Galactic latitudes (caused by the assumption of an infinitely extended disc), we apply the distance prior corresponding to stars at $b = 1^{\circ}$ to all the cases in the north Galactic hemisphere, and that of stars at $b = -1^{\circ}$ to those in the south Galactic hemisphere.

We calculate the median, mode, and mean of the resulting posterior distance distribution of each star, and estimate the uncertainty by calculating the $16$th and $84$th percentiles.

Even if our prior is simpler than that of \citet{Bailetal21} (as there is only a dependence on Galactic latitude), and its parameters were fitted to \textit{Hipparcos} data, the resulting distances have still better performance for OB stars (see the discussion in section \ref{subsec:noBJ}). Furthermore, the estimated distances are robust even under significant variations in the model parameters, suggesting that the results are driven more by the data than by the prior (making the need for a more complex prior distribution based on \textit{Gaia} data largely unimportant).

\subsubsection{Astrometry outside of \textit{Gaia} DR3}
\label{subsec:usno}

Some of the most important OB stars lack \textit{Gaia} DR3 counterpart, or when they do have counterparts, the relevant astrometry is missing. This may be due to several factors: \textbf{1)} some stars are too bright, causing saturation of the camera CCD (e.g., $\alpha$ Vir $=$ Spica $=$ GLS 14 813; with G = 0.9 mag, or $\kappa$ Ori $=$ Saiph $=$ GLS 14 795; with G = 2.1 mag), \textbf{2)} some are too faint, falling below the sensitivity threshold  (e.g., Sgr A$^{\star}$ S02 $=$ GLS 20 981 or WR 63 $=$ GLS 3289); and \textbf{3)} others are too close to another \textit{Gaia} DR3 source, meaning that they are difficult to separate or have contrasts that are too extreme (e.g. $\tau$ CMa E $=$ GLS 14 805 E or MY Cam B $=$ GLS 7836 B). Because of this, there are \unum{\num{615}} stars without \textit{Gaia} DR3 astrometry in the ALS III.

Given the high interest in distances and proper motions for many of these stars, we address this issue in several ways. For the bright stars lacking \textit{Gaia} DR3 astrometry, we rely on the USNO Bright Star Catalogue (UBSC; \citealt{Zachetal22}), retrieving parallaxes (which are “\textit{of limited value because they closely follow the Hipparcos measures}”), calculating distances (with the same procedure as in section \ref{subsec:dist} but without any corrections for the parallaxes), and proper motions (which are significantly improved over those measured by Hipparcos). For close sources, we either inherit the astrometry from the companion or, when applicable, use the average astrometry of the cluster. To distinguish between these cases, we introduce the "Aflag" column in the tables. This column takes the following values: "G" when the astrometry is from \textit{Gaia} DR3 (\unum{\num{19772}} cases), "U" when the astrometry comes from the UBSC-Hipparcos (\unum{\num{61}} cases), "C" when the astrometry is inherited from a close companion with available astrometry or averaged from the host cluster (in these \unum{\num{56}} cases we can use the recipe for obtaining group parallaxes in \citet{Maizetal21c}, which takes the angular covariance into account), and "N" when no astrometry is available (\unum{\num{919}} cases). When non-\textit{Gaia} distances are retrieved from the literature, without any parallax measurement present, the value is assigned to the ``Dist\_P50'' column (see Table~\ref{tab:cat_cols} for more information). In ALS III, we have not yet implemented the use of group distances for individual stars from the Villafranca project (see above), but this may be included in future versions.

\subsection{Structure of the catalogue}
\label{subsec:structcat}

\begin{table*}
    \centering
    \vspace{-5 pt}
    \begin{tabular}{l c c c l}
        \hline
        Column name & Number & Note & Units & \multicolumn{1}{c}{Description} \\
        \hline
        \multicolumn{5}{c}{\textbf{Catalogue categories, identifiers and notes}}\\
        Cat & \unum{\num{20808}} & a & - & Category / internal catalogue partition (M/I/L/H/E/A/C/D/X/U; see Sect. \ref{subsec:structcat})\\
        Pflag & \unum{\num{20808}} & a & - & GLS/ALS prefix assignment channel (A1/G2/G3/G4/G5/A6/G6/A7/G7/A8/G8; see Sect. \ref{subsec:prefix})\\
        ID\_ALS & \unum{\num{20808}} & a & - & Alma Luminous Star identifier \\
        Name & \unum{\num{20387}} & b & - & Principal name for the star \\
        ID\_GOSC & \unum{\num{20387}} & b & - & Galactic O-Star Catalog identifier \\
        ID\_GWRC & \unum{\num{668}} & - & - & Galactic Wolf Rayet Catalogue identifier \\
        ID\_DR3 & \unum{\num{20018}} & - & - & \textit{Gaia} DR3 source identifier \\
        ID\_2MASS & \unum{\num{19983}} & - & - & 2MASS Point Source Catalogue identifier \\
        ID\_SIMBAD & \unum{\num{20272}} & - & - & Main Simbad identifier \\
        ID\_OTHER & \unum{\num{15010}} & - & - & Additional common identifiers (separated by ;)\\
        ID\_DUP & \unum{\num{403}} & - & - & ALS identifier of the principal star, if this is a catalogue duplicate\\
        ID\_BIN & \unum{\num{106}} & - & - & ALS identifier of the companion, if the ``GLS'' prefix was inherited from it (see Sect. \ref{subsec:prefix})\\
        Comments & \unum{\num{1020}} & - & - & Notes\\

        \multicolumn{5}{c}{\textbf{Coordinates}}\\
        RA & \unum{\num{20387}} & b & hh:mm:ss & Right ascension (at epoch 2016.0)\\
        DEC & \unum{\num{20387}} & b & dd:mm:ss & Declination (at epoch 2016.0)\\
        GLON & \unum{\num{20387}} & b & deg & Galactic longitude\\
        GLAT & \unum{\num{20387}} & b & deg & Galactic latitude\\

        \noalign{\smallskip}
        \multicolumn{5}{c}{\textbf{Photometry}}\\
        Gmag & \unum{\num{20010}} & - & mag & $G$ magnitude in \textit{Gaia} DR3\\
        Gmag\_cor & \unum{\num{20010}} & - & mag & $G'$; corrected $G$ magnitude (see Sect. \ref{subsec:calibphot})\\
        Gmag\_error & \unum{\num{20010}} & - & mag & $\sigma_{\text{G}}$; $G/G'$ magnitude uncertainty (see Sect. \ref{subsec:calibphot})\\
        Gabs & \unum{\num{19825}} & - & mag & Absolute $G'$ magnitude, from the median of the posterior for the distance\\
        Gabs\_error & \unum{\num{19825}} & - & mag & Absolute $G'$ magnitude uncertainty, from the 16th and 84th percentiles of the posterior\\
        BPmag & \unum{\num{19955}} & - & mag & $G_{\text{BP}}$ magnitude in \textit{Gaia} DR3\\
        BPmag\_error & \unum{\num{19955}} & - & mag & $\sigma_{\text{BP}}$; $G_{\text{BP}}$ magnitude uncertainty\\
        RPmag & \unum{\num{19961}} & - & mag & $G_{\text{RP}}$ magnitude in \textit{Gaia} DR3\\
        RPmag\_error & \unum{\num{19961}} & - & mag & $\sigma_{\text{RP}}$; $G_{\text{RP}}$ magnitude uncertainty\\
        Cstar & \unum{\num{19955}} & - & mag & $C^*$: Corrected $G_{\text{BP}}-G$ and $G-G_{\text{RP}}$ flux excess factor from \citet{Rieletal21}\\
        Jmag & \unum{\num{19983}} & - & mag & $J$ magnitude in 2MASS\\
        Jmag\_error & \unum{\num{19667}} & - & mag & $\sigma_J$; $J$ magnitude uncertainty in 2MASS\\
        Hmag & \unum{\num{19908}} & - & mag & $H$ magnitude in 2MASS\\
        Hmag\_error & \unum{\num{19831}} & - & mag & $\sigma_H$; $H$ magnitude uncertainty in 2MASS\\
        Kmag & \unum{\num{19978}} & - & mag & $K$ magnitude in 2MASS\\
        Kmag\_error & \unum{\num{19840}} & - & mag & $\sigma_K$; $K$ magnitude uncertainty in 2MASS\\
        Qflag & \unum{\num{19983}} & - & - & $JHK$ photometric quality flag from 2MASS\\

        \noalign{\smallskip}
        \multicolumn{5}{c}{\textbf{Parallax and distances}}\\
        Plx & \unum{\num{19889}} & c & mas & $\omega_c$; corrected \textit{Gaia} DR3 parallax$^{\textit{e}}$ (see Sect. \ref{subsec:parallax})\\
        Plx\_error & \unum{\num{19889}} & c & mas & $\sigma_{\text{ext}}$; corrected \textit{Gaia} DR3 parallax uncertainty$^{\textit{e}}$ (see Sect. \ref{subsec:parallax})\\
        Dist\_mean & \unum{\num{19889}} & c & pc & Mean of the posterior distribution for the distance$^{\textit{e}}$ (see Sect. \ref{subsec:dist})\\
        Dist\_mode & \unum{\num{19889}} & c & pc & Mode of the posterior distribution for the distance$^{\textit{e}}$ (see Sect. \ref{subsec:dist})\\
        Dist\_median & \unum{\num{19889}} & c & pc & $d_{\text{OB}}$; median of the posterior distribution for the distance$^{\textit{e}}$ (see Sect. \ref{subsec:dist})\\
        Dist\_errorm & \unum{\num{19889}} & c & pc & Lower distance uncertainty: median minus the 16th percentile of the posterior distribution$^{\textit{e}}$ (see Sect. \ref{subsec:dist})\\
        Dist\_errorp & \unum{\num{19889}} & c & pc & Upper distance uncertainty: 84th percentile minus the median of the posterior distribution$^{\textit{e}}$ (see Sect. \ref{subsec:dist})\\
        Aflag & \unum{\num{20808}} & a & - & Astrometric reference; (G/U/C/N; see Sect. \ref{subsec:usno})\\
        RUWE & \unum{\num{19772}} & d & - & Renormalized Unit Weight Error in \textit{Gaia} DR3\\

        \noalign{\smallskip}
        \multicolumn{5}{c}{\textbf{Proper motions}}\\
        PM\_RA & \unum{\num{19889}} & c & mas/a & Corrected \textit{Gaia} DR3 proper motion component in right ascension$^{\textit{e}}$ (see Sect. \ref{subsec:pm})\\
        PM\_RA\_error & \unum{\num{19889}} & c & mas/a & Corrected \textit{Gaia} DR3 proper motion uncertainty in right ascension$^{\textit{e}}$ (see Sect. \ref{subsec:pm})\\
        PM\_DEC & \unum{\num{19889}} & c & mas/a & Corrected \textit{Gaia} DR3 proper motion component in declination$^{\textit{e}}$ (see Sect. \ref{subsec:pm})\\
        PM\_DEC\_error & \unum{\num{19889}} & c & mas/a & Corrected \textit{Gaia} DR3 proper motion uncertainty in declination$^{\textit{e}}$ (see Sect. \ref{subsec:pm})\\
        PM\_GLON & \unum{\num{19889}} & c & mas/a & Proper motion component in Galactic longitude$^{\textit{e}}$ (see Sect. \ref{subsec:pm})\\
        PM\_GLON\_error & \unum{\num{19889}} & c & mas/a & Proper motion uncertainty in Galactic longitude$^{\textit{e}}$ (see Sect. \ref{subsec:pm})\\
        PM\_GLAT & \unum{\num{19889}} & c & mas/a & Proper motion component in Galactic latitude$^{\textit{e}}$ (see Sect. \ref{subsec:pm})\\
        PM\_GLAT\_error & \unum{\num{19889}} & c & mas/a & Proper motion uncertainty in Galactic latitude$^{\textit{e}}$ (see Sect. \ref{subsec:pm})\\

        \noalign{\vskip 6pt}
        \multicolumn{5}{l}{\parbox{0.85\textwidth}{$^{\textit{a}}$ The number of total the entries in the catalogue.}}\\
        \multicolumn{5}{l}{\parbox{0.85\textwidth}{$^{\textit{b}}$ The number of entries in the catalogue that are not duplicates (Cat = "D") nor lost stars (Cat = "X").}}\\
        \multicolumn{5}{l}{\parbox{0.85\textwidth}{$^{\textit{c}}$ Stars having astrometric data either from \textit{Gaia} DR3, UBSC or inherited from a companion having astrometry (see Aflag).}}\\
        \multicolumn{5}{l}{\parbox{0.85\textwidth}{$^{\textit{d}}$ Stars having astrometric data from \textit{Gaia} DR3.}}\\
        \noalign{\vskip 6pt}
        \multicolumn{5}{l}{\parbox{0.85\textwidth}{$^{\textit{e}}$Unless Aflag is "U" or "C", in which case the values for the astrometry are taken directly from the UBSC or inherited from a companion, respectively (see Sect. \ref{subsec:usno}).}}\\
        \hline
    \end{tabular}
    \caption{Description for the columns in the ALS III catalogue, as sent to the CDS.}
    \label{tab:cat_cols}
\end{table*}

The ALS III catalogue is divided into 10 internal subcatalogs, following a methodology similar to that described in Paper II. Five of these subcatalogs (D, X, U, C, and A) represent the residuals from the data reduction process, which identifies high-quality astrometric and photometric stars from \textit{Gaia} DR3. The resulting sample of reliable stars is further subdivided into five additional subcatalogs (M, I, L, H, and E) based on astro-photometric selection criteria. The left branch of the decision tree in Figure \ref{fig:flowchart} provides a hierarchical overview of the steps involved in these partitions.\footnote{To avoid confusions arising from the names assigned to each internal catalogue, it is important to remember that, in the context of this work, M-stars and A-stars are not M- and A-type stars, nor L-stars are L-type brown dwarfs, nor I-stars are to be seen as supergiants.}

\begin{figure}
    \centering
    \includegraphics[width=0.47\textwidth, trim = 14 0 0 0, clip]{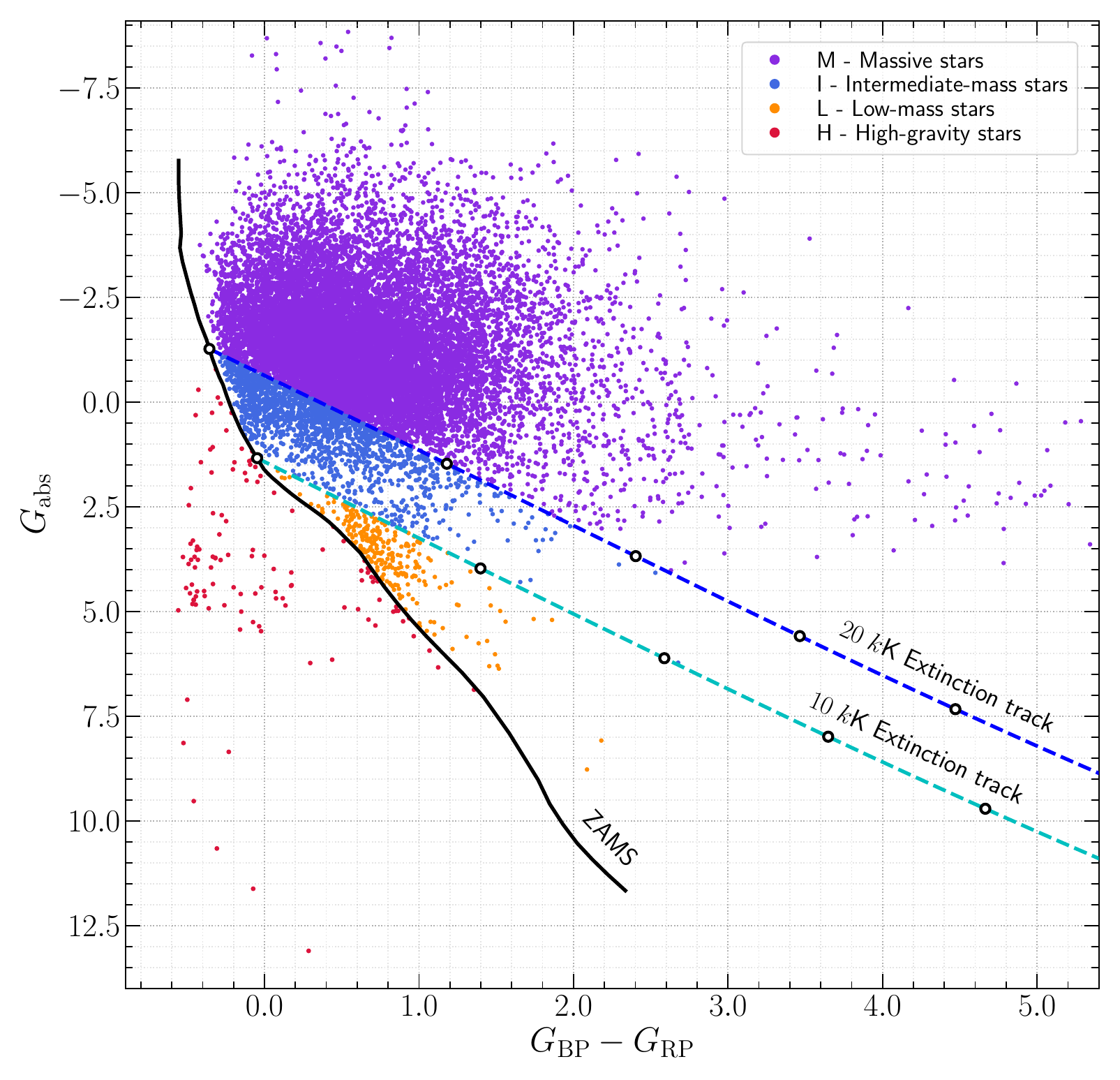}
    \caption{\textit{Gaia} DR3 color-absolute magnitude diagram, with the M, I, L and H samples (colour-coded), selected by slicing by the ZAMS, the $10$ kK and $20$ kK extinction tracks. Markings on the extinction curves correspond to $E(4405-5495) = 0$, $1$, $2$, $3$, and $4$ mag. The gap between the ZAMS and highly luminous stars can be partially attributed to the color and brightness shift caused by extinction, but not entirely (\citealt{Holgetal20}).}
    \label{fig:HR}
\end{figure}

\subsubsection{Data reduction}
\label{subsec:datared}
Starting from the whole catalogue, we sequentially reduce it by applying different criteria and putting the excluded entries in each of these categories:

\begin{itemize}
    \item \textbf{D} contains the duplicates of the catalogue. These are entries in the original ALS and the ALS (P) that are redundant, but are still preserved in the ALS III for historical reasons.
    \item \textbf{X} contains entries of the catalogue that are currently meaningless or do not reflect real stars. \unum{11} of these are simply ``lost stars''; entries that appear to reference photographic plates where the actual objects are nowhere to found and the coordinates are either wrong or plainly missing from any reference. Stars ALS 2275 and ALS \num{19500} are obsolete designations for several components of the NGC 3603 cluster's core. The entry for ALS 4340 actually refers to the nova event of 1968, and there are some additional ambiguous references.
    \item \textbf{U} contains stars that have no \textit{Gaia} DR3 counterpart. The vast majority of these have validated spectral classifications and include well known OB stars in highly obscured regions (too faint for \textit{Gaia}) and a few nearby OB stars that are too bright for \textit{Gaia}. Without \textit{Gaia} DR3 data we can't use the color-absolute magnitude diagram to photometrically divide the sample. All the entries in this internal catalogue have been examined manually and will have additional comments in the ALS web page for future reference.
    \item \textbf{A} contains stars with \textit{Gaia} DR3 counterparts and good-quality photometry that meet at least one of the following criteria: 1) lack parallax data, or 2) have a relative uncertainty in the parallax larger than $1/3$, or 3) have $RUWE$ values larger than $1.4$. Some \unum{\num{1006}} of these stars are ``GLS'' stars, which implies that as future \textit{Gaia} data releases improve astrometry, it is likely that the number of stars in the M internal catalogue originating from this category will increase, potentially adding more OB stars to the 3D Galactic map.
    \item \textbf{C} contains stars with \textit{Gaia} DR3 counterparts that either 1) lack photometry in $G$, $G_{\text{BP}}$, or $G_{\text{RP}}$, or 2) have a corrected $G_{\text{BP}}-G$ and $G-G_{\text{RP}}$ flux excess factor, $C^*$ (\citealt{Rieletal21}), outside the $[-0.10 \: ;\:0.15]$ range. Such colour deviations can result from nebular contamination, crowded fields, unresolved binarity, or general inaccuracies in the photometric data.
\end{itemize}

As expected, duplicate entries (D) and lost stars (X), all follow the "A8" path for the prefix assignment and lack data in several fields of the catalogue, as they are maintained only to preserve their identifiers and allow traceability in the literature. Ideally, if the catalogue were being constructed entirely from scratch, these entries, and their associated identifiers, would have been removed.

\subsubsection{Astrophotometric catalogue partition}
\label{subsec:cats}

Catalogue entries that are not in the D, X, U, C and A categories are stars having enough \textit{Gaia} DR3 good-quality data to be reliably placed in the color-absolute magnitude diagram (figure \ref{fig:HR}). We use the ZAMS curve from the solar metallicity grid in \citet{Maiz13a}, and the extinction tracks for ZAMS stars with $10$ kK and $20$ kK, following the extinction law from \citet{Maizetal14a} (and using a $R_{5495} = 3.0$ from \citet{MaizBarb18}, which is typical for OB stars at intermediate-to-high extinctions) as boundaries in the color-absolute magnitude diagram from which to subdivided the sample even further into these photometric classes:

\begin{itemize}
    \item \textbf{M}, contains stars above the ZAMS and the $20$ kK extinction track. We consider these to be likely massive stars according to \textit{Gaia} DR3 photometry, and conditioned by the fact that as part of the ALS they have been reported as OB stars in the literature or even have validated spectral classifications.
    \item \textbf{I}, contains stars between the $10$ kK and 20 $kK$ extinction tracks. These are likely intermediate-mass stars. Some of these were previously misidentified as OB stars in the literature, while others might eventually be promoted to the M category on future \textit{Gaia} date releases, as they represent OB stars near the limits of our current classification criteria (something supported by the \unum{$88$} members of this category that have GOSC or GWRC validation as OB stars, by following the G2 or G3 paths of the flow chart \ref{fig:flowchart}).
    \item \textbf{L}, contains stars that are located above the ZAMS but below the $10$ kK extinction track, which implies that they are probably low-mass stars. This classification is supported by all GOSC-validated spectra for stars in this category, which consistently confirm their low-mass nature. The only GLS star in this category is GLS \num{15161}, which is considered to be a component of the Cyg OB2-22 system that still lacks spectroscopic classification (thus inheriting the prefix from the companions).
    \item \textbf{H}, contains stars below and to the left of the ZAMS, which implies high-gravity objects like hot white dwarfs and hot sub-dwarfs. The only GLS stars in this category is GLS \num{22395}, a validated B2: star that may present some issues in its \textit{Gaia} DR3 photometry.
    \item \textbf{E}, contains seven extragalactic stars (six from the LMC and one from the SMC) that will remain in the ALS III for historical reasons. The AA Doradus = ALS \num{16654} system, a well-known hot sub-dwarf binary within our Galaxy, coincidentally lies along the line of sight of the central field of the LMC, and as a result, it is retained in the H category rather than being included here.
\end{itemize}

In sumary, to assess the reliability of entries in the ALS III as OB stars, readers can use two key criteria; a photometric one (the star is listed in the M internal catalogue), and a spectroscopic one (the star having a ``GLS'' prefix and in particular a ``Pflag'' of G2, G3, or G5, as described in section \ref{subsec:prefix}). The self-consistency of the catalogue is apparent when we consider that \unum{$98.9\%$} of the M stars have a ``GLS'' prefix and \unum{$90.2\%$} of the stars with a ``GLS'' prefix are members of the M catalogue (as seen in the confusion matrix in Figure \ref{fig:flowchart}).

\section{Results}

In this section we present three analyses that can be immediately carried out with ALS III. The first one is a comparison among \textit{Gaia}-based distances from different sources. The second one is how sample selection affects the determination of parameters determined from \textit{Gaia} data. The third one is how ALS III improves our knowledge of the distribution of OB stars in our immediate surroundings.

\subsection{Comparing distance estimates}
\subsubsection{Bailer-Jones distances}
\label{subsec:noBJ}

\begin{figure} 
    \centering
    \includegraphics[width=0.48\textwidth, trim = 14 14 0 5, clip]{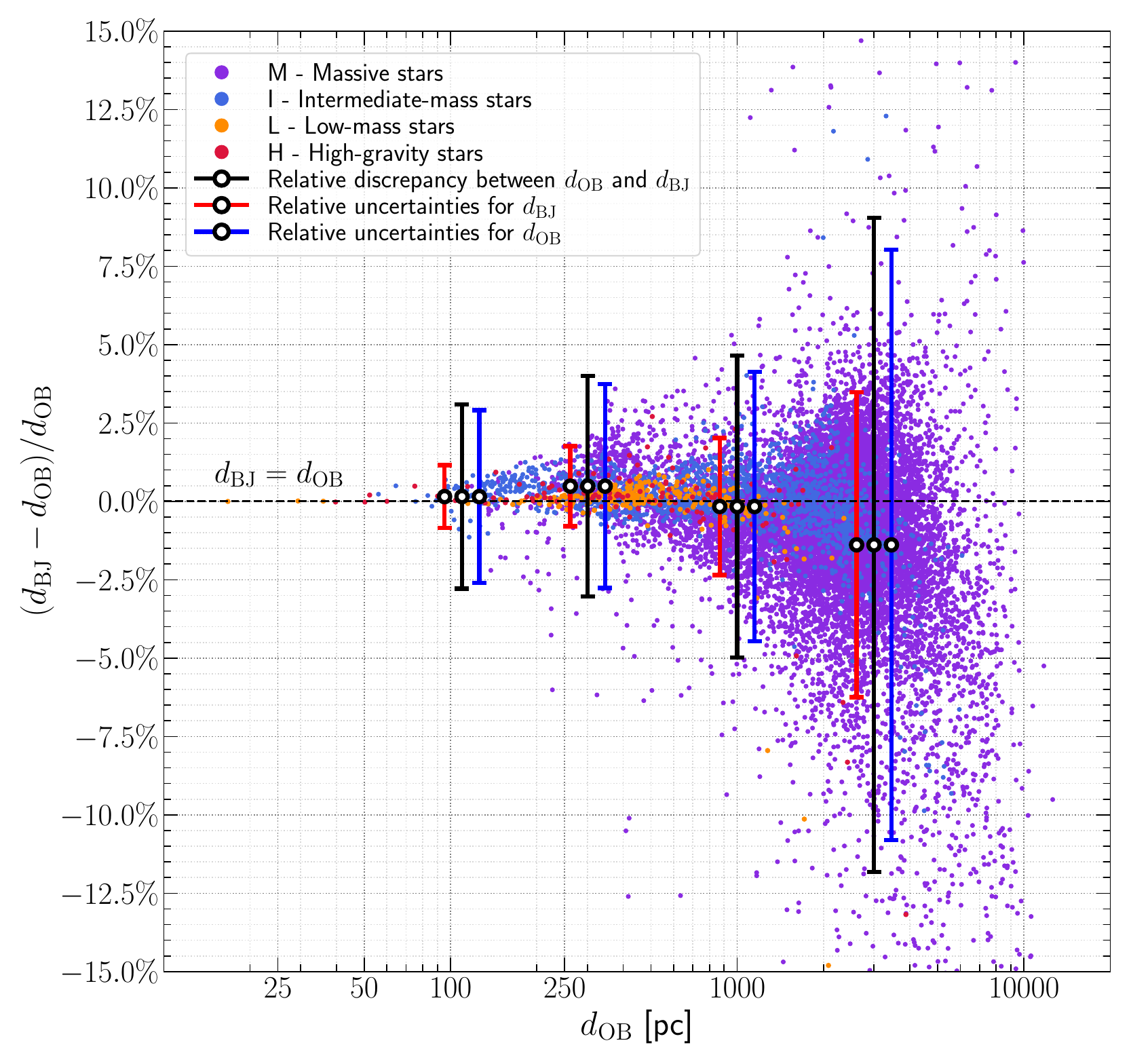}
    \vspace{-3pt}
    \caption{Relative discrepancy between the distance estimates based on the OB prior (this work) and those using the general prior in \citet{Bailetal21}. The comparison is made between the medians of the posterior distributions. The error bars in black show one standard deviation from the relative discrepancy between distances for typical stars at $0.1$, $0.3$, $1$, and $3$ kpc, respectively. The error bars in red (shifted to the left of the black error bars, for better visibility) and the error bars in blue (shifted to the right of the black error bars) span the relative uncertainties in the $d_{\text{BJ}}$ and $d_{\text{OB}}$ estimates, respectively.}
    \label{fig:dist}
\end{figure}

Our distance estimates, $d_{\text{OB}}$, as well as those from \citet{Bailetal21}, $d_{\text{BJ}}$, rely on a Bayesian approach to address common biases associated with the naive calculation of distance as the inverse of the parallax. Our prior is specifically tailored for the young, massive stellar population, while the prior used by \citet{Bailetal21} is designed for a broader (and thus late-type dominated) \textit{Gaia} population, as we described in section \ref{subsec:dist}.

In Figure \ref{fig:dist}, we compare both distance estimates by examining the relative discrepancy between them as a function of distance. The error bars represent the typical uncertainty associated with the posterior distributions of our estimates. This comparison reveals that our distance estimates are generally in agreement (within the error margins) with those of \citet{Bailetal21} for distances beyond ${\sim} 1$ kpc. In general for stars below ${\sim} 1$ kpc the discrepancies are contained at around $2.5\%$, with \citet{Bailetal21} systematically overestimating them the first ${\sim}0.5$ kpc and underestimating them at larger distances.

More importantly, it is clear from Figure \ref{fig:dist} that the relative uncertainties in $d_{\text{BJ}}$ (in red) are underestimated by a factor of $\sim 2$ when compared to the relative uncertainties in our distance estimates (in blue). This is expected as \citet{Bailetal21} did not apply corrections to the parallax uncertainties, erroneously yielding a higher confidence.

\subsubsection{ALS II distances}

We can now compare the distances obtained here with those presented in Paper II, where \textit{Gaia} DR2 was used. To facilitate this comparison, we have included markers at specific distances in the upper right panel of Figure \ref{fig:Maps}, along with average typical error bars derived from the posterior distributions of stars at those distances (similar to figure 5 in Paper II). From this, we can observe that the spread of OB associations in the radial direction is on a similar scale to the typical distance uncertainties. This suggests that with \textit{Gaia} DR4, we are likely to see the spiral arms as even narrower structures, as the uncertainties in parallax are expected to further decrease.

From these numbers we can tell that the relative uncertainty in distance went from $5.2\%$ in ALS II - \textit{Gaia} DR2 to $3.6\%$ in the ALS III - \textit{Gaia} DR3 for stars at $1$ kpc. For stars at $2$ kpc it decreased from $9.0\%$ to $7.1\%$, and for stars at $3$ kpc it went from $13.6\%$ to $8.4\%$, making the structures of the map believable up to ${\sim}3$ kpc. We can compare those uncertainties with the ones derived for stellar groups in Villafranca~II \citep{Maizetal22a}, which can be approximately characterized by the distance in kpc expressed in percentage e.g. 1\% at 1~kpc or 3\% at 3~kpc. They are significantly lower (by a factor of $\sim$3 for most of the range of interest) because \textit{Gaia}~DR3 group distance uncertainties are dominated by systematic effects (the angular covariance term in \citealt{Maizetal21c}) while the corresponding individual distance uncertainties are dominated by the random uncertainties (the first term in Eqn.~\ref{eq:extunc}).

\subsection{Validation of \textit{Gaia} DR3 temperature and extinction determinations for massive stars}

\begin{figure*}
    \centering
    \includegraphics[width=0.587\textwidth, trim=5 5 0 5, clip]{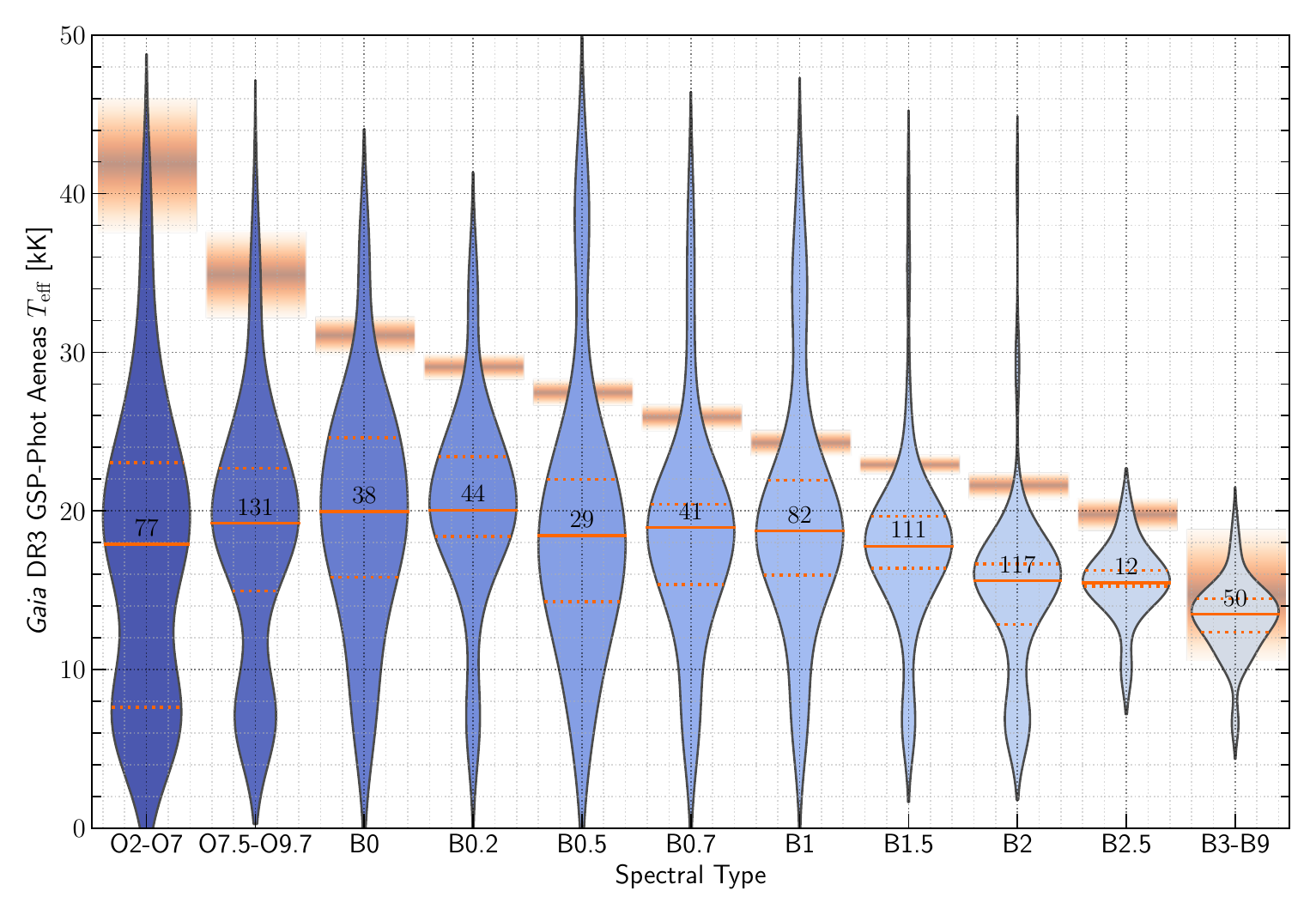}
    \includegraphics[width=0.40\textwidth, trim=0 5 5 5, clip]{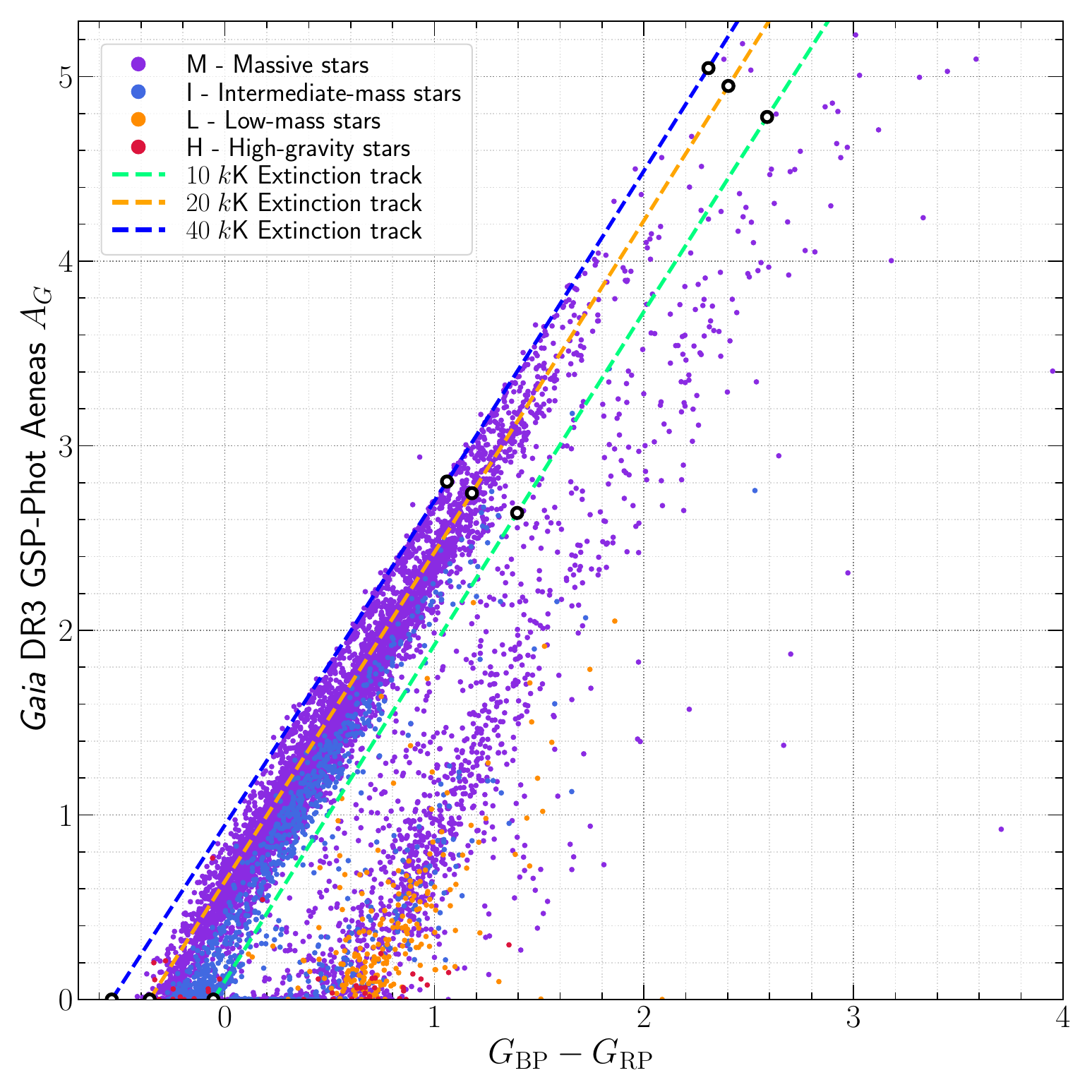}
    \caption{(left) Violin plot of \textit{Gaia} DR3 GSP-Phot Aeneas \Teff\ given as smoothed distributions for the verified spectral types of MS massive stars in the ALS III. The numbers inside the violins indicate the size of each sample, the orange lines mark the median (continuous), and the first and third quartiles (dotted) for each sample. The orange gradients span the actual temperatures for each subtype as reported in \citet{Holgetal18}, \citet{Trunetal07} and \citet{PecaMama13}. (right) Broadband $A_G$ extinction estimates from \textit{Gaia} DR3 GSP-Phot Aeneas against the $G_{\text{BP}}-G_{\text{RP}}$ colour. We also plot the expected relationship between the two quantities for MS stars of $10$, $20$, and $40$ $k$K and a extinction law from \citet{Maizetal14a}, with $R_{5495} = 3.0$. Markings on the extinction curves correspond to $E(4405-5495) = 0$, $1$, and $2$ mag.}
    \label{fig:violin_GSPPHOT}
\end{figure*}

\begin{figure*}
    \centering
    \includegraphics[width=0.587\textwidth, trim=5 5 0 5, clip]{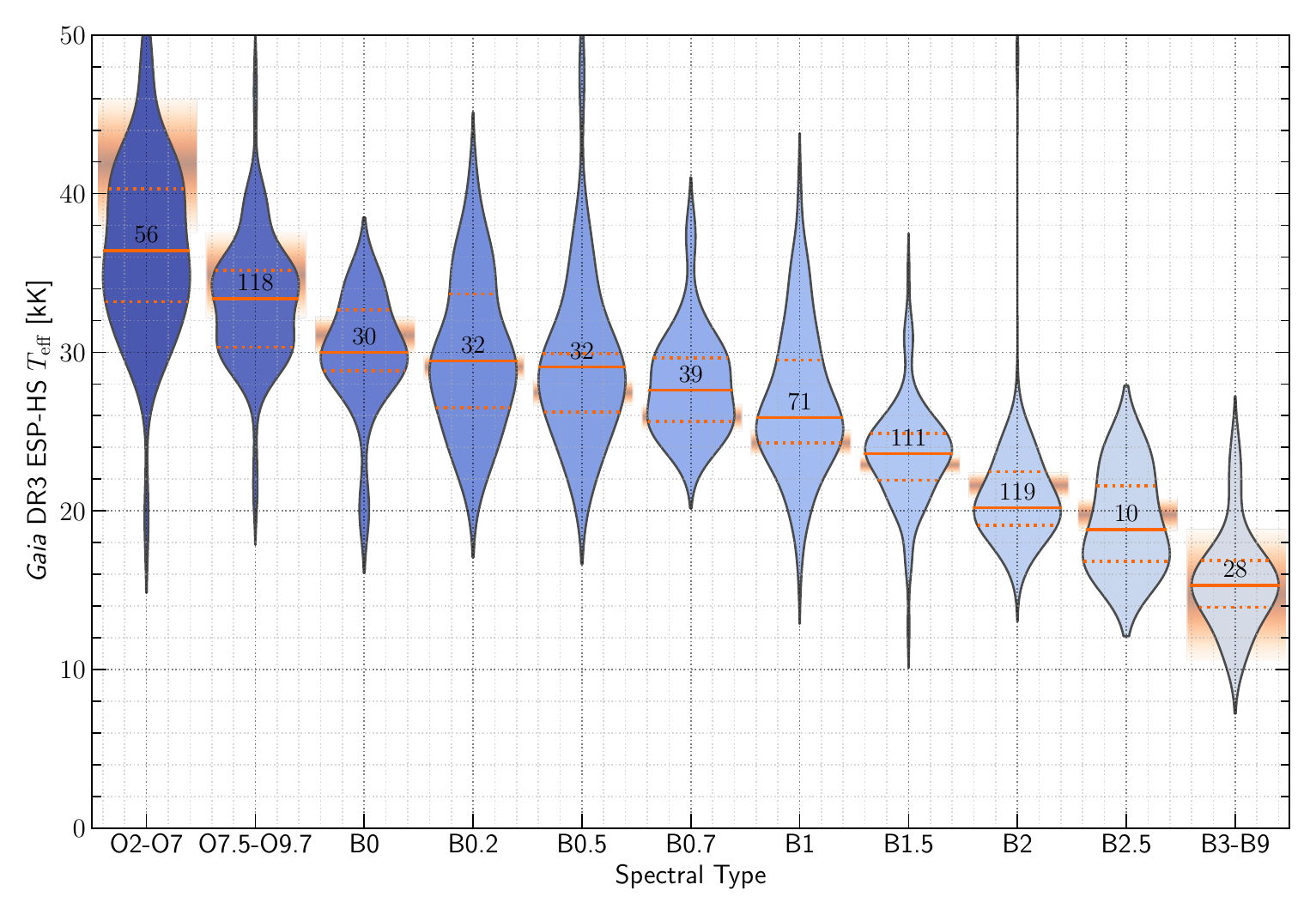}
    \includegraphics[width=0.40\textwidth, trim=0 5 5 5, clip]{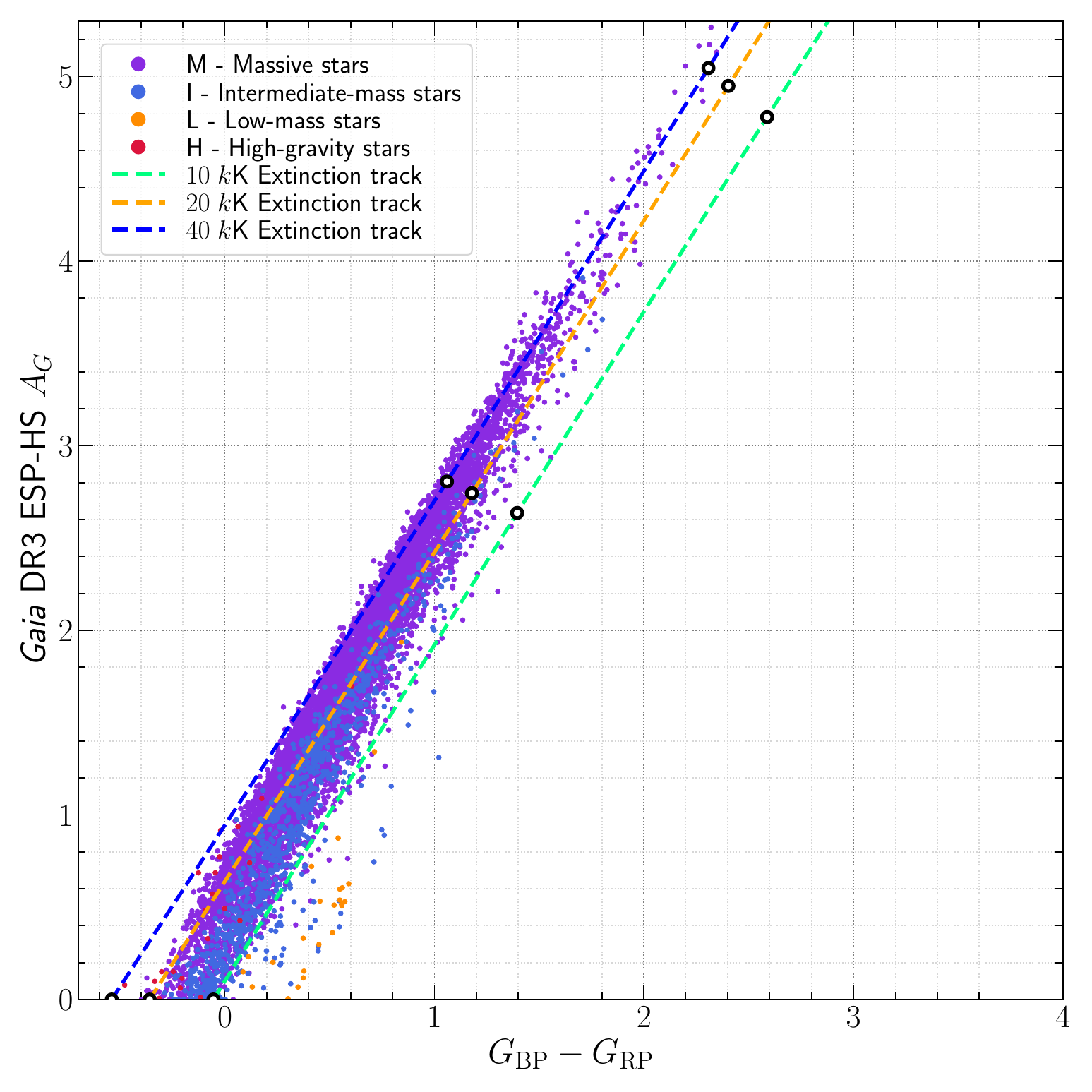}
    \caption{Same as in figure \ref{fig:violin_GSPPHOT} but using the \textit{Gaia} DR3 ESP-HS pipeline for \Teff\ (left panel) and  $A_G$ extinction (right panel) instead of the values derived by GSP-Phot Aeneas.}
    \label{fig:violin_ESPHS}
\end{figure*}

\textit{Gaia} DR3 provides estimates for some astrophysical parameters for several stars, including $T_{\text{eff}}$ and broadband $G$ extinctions, $A_G$, by means of the ``General Stellar Parametrizer from Photometry'' (GSP-Phot) and also from the ``Extended Stellar Parametrizer for Hot Stars'' (ESP-HS). Both pipelines are part of the Apsis module in the \textit{Gaia} data processing system (\citealt{Creeetal23a, Foueetal23}), with the ESP-HS considered to be specifically well suited for hot stars.

\citet{Frem24} recently examined the quality of the Gaia DR3 Apsis modules. By comparing the data with SIMBAD and spectroscopic surveys such as LAMOST and the Gaia-ESO Survey, they found that the classification of ``hot'' stars (actually OBA) in the Apsis modules can be largely compromised by contamination from cooler stars, such as white dwarfs, subdwarfs, and RR Lyrae stars. They also highlight that O-type classifications are inaccurate, with frequent confusion with B-type stars. Both GSP-Phot and ESP-HS tend to underestimate the effective temperatures of massive stars due to the degeneracy between temperature and interstellar extinction, leading to many M-type stars been incorrectly marked as highly reddened O-type stars.

One of the main applications of our catalogue is to independently validate or challenge the accuracy of these parameters when dealing with OB stars. In figure \ref{fig:violin_GSPPHOT}, we use the $T_{\text{eff}}$ and $A_G$ solutions from GSP-Phot, and in figure \ref{fig:violin_ESPHS} the ones from the ESP-HS. In the left panels of figures \ref{fig:violin_GSPPHOT} and \ref{fig:violin_ESPHS}, we select only spectroscopically validated main sequence stars in the ALS III with relevant \textit{Gaia} data. We then plot the \Teff\ distributions for each pipeline, grouped by different subtypes. The ranges for the subtypes are chosen to ensure a sufficient number of stars for statistical analysis. Known temperature ranges for each subtype grouping are taken from \citet{Holgetal18} (using the empirical law for O-type dwarfs), \citet{Trunetal07} (for B0 to B2.5 dwarfs) and \citet{PecaMama13} (for B3 to B9 dwarfs), for better comparison. 

For GSP-Phot, the variance in the \Teff\ distributions increases as we go towards earlier spectral types, to the point that O2-O7 stars are said to have solar $T_{\text{eff}}$. Furthermore, the median of the temperatures are almost incapable of reaching the $20$ kK of the coldest OB stars. The discrepancy between known typical ranges for temperatures for each subtype and the ones given by GSP-Phot make clear that this pipeline is extremely unreliable for estimating temperatures of OB stars (as expected). The only accurate match is obtained for B3-B9 main sequence stars, which are not massive stars. For the rest, the values of \Teff\ are underestimated.

Temperature results are better for ESP-HS as, in general, the medians of the temperatures are close to the expected ones. The significant exception is the first bin (O2-O7), where ESP-HS underestimates the values of \Teff. However, the variances are still too large.

In regards to extinction, the GSP-Phot results (right panel in Fig.~\ref{fig:violin_GSPPHOT}) deviate from the expected trend (between the dashed orange and blue lines) in two ways: First, the \AGG\ values are underestimated, especially for low extinction. This is likely associated with the global \Teff\ underestimation previously indicated. Second, a different, quasi-parallel trend is found to the right of the main trend. Those objects have not only their extinction severely underestimated but they are also identified as cool stars. In summary, GSP-Phot slightly underestimates \AGG\ for most of the sample and does it to a greater degree for a small subsample.

The ESP-HS extinction results (right panel in Fig.~\ref{fig:violin_ESPHS}) are significantly better. The second, parallel trend disappears and the main trend is better located within the dashed orange and blue lines. We note, however, that \AGG\ values are still underestimated in the low extinction regime.

%(right panels in figures \ref{fig:violin_GSPPHOT} and \ref{fig:violin_ESPHS}), both pipelines have improved significantly in the estimates given by Apsis in \textit{Gaia} DR2, which we examined in Paper II (left panel in figure 3) in the same way we are doing here. In Paper II the estimated extinction had an abrupt discontinuity around $G_{\text{BP}}-G_{\text{RP}} = 1.1$ mag basically resetting $A_G$ to zero for redder stars. Something similar, albeit more complicated, can be seen for the $A_G$ derived by GSP-Phot. On the other hand, ESP-HS values are in good agreement with the extinction expected when following the law in \citet{Maizetal14a}, with $R_{5495} = 3.0$. The only major discrepancy comes from the inclination of the extinction against colour.

In summary, the GSP-Phot \Teff\ and extinction results are unreliable in general. The equivalent ESP-HS results are significantly better but still show a small \AGG\ bias at low extinction and a significant dispersion in \Teff.

\subsection{An improved view of the solar neighbourhood with OB stars}
\label{subsec:maps}

\begin{figure*}
    \centering
    \hspace{-0.9cm}
    \begin{minipage}{0.48\textwidth}
        \centering
        \includegraphics[width=\textwidth, trim=20 5 0 0, clip]{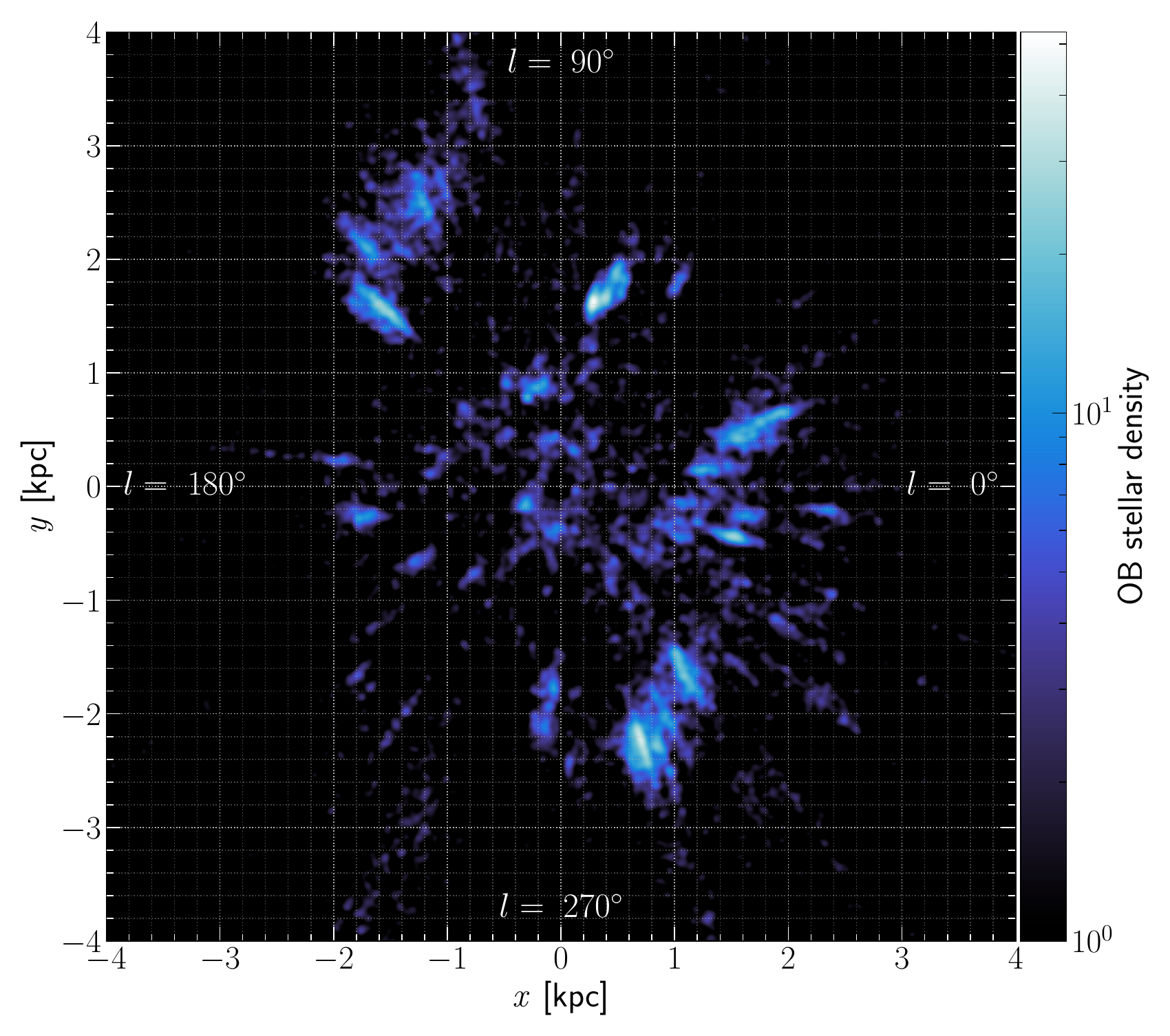}
    \end{minipage}
    \hspace{0.4cm}
    \begin{minipage}{0.43\textwidth}
        \centering
        \includegraphics[width=\textwidth, trim=10 5 0 0, clip]{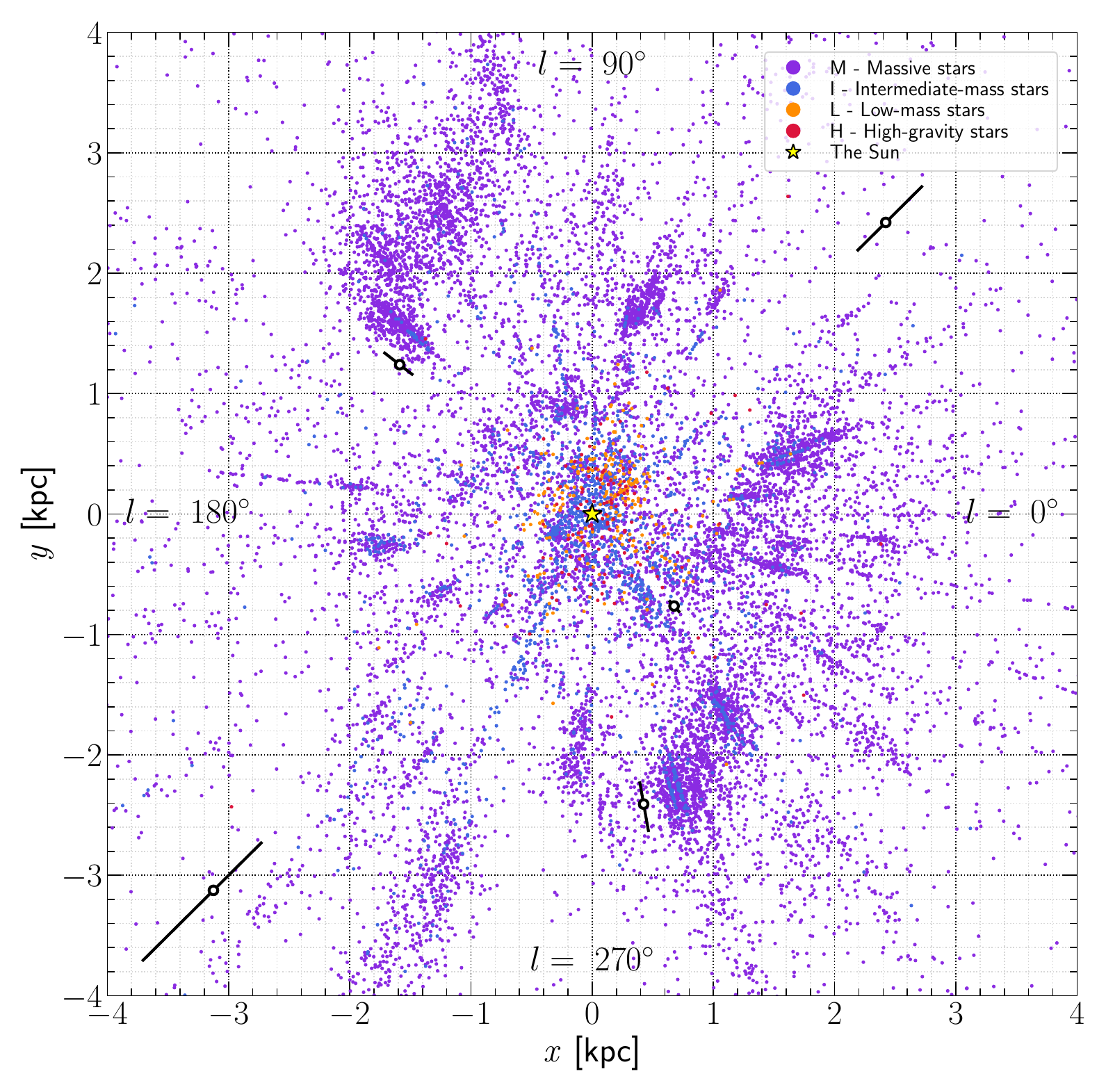}
    \end{minipage}
    
    \vspace{-0.47cm}  % Add vertical space between rows
    \hspace{0.1cm}
    
    % Second row of subfigures
    \begin{minipage}{0.485\textwidth}
        \centering
        \includegraphics[width=\textwidth, trim=13 5 0 0, clip]{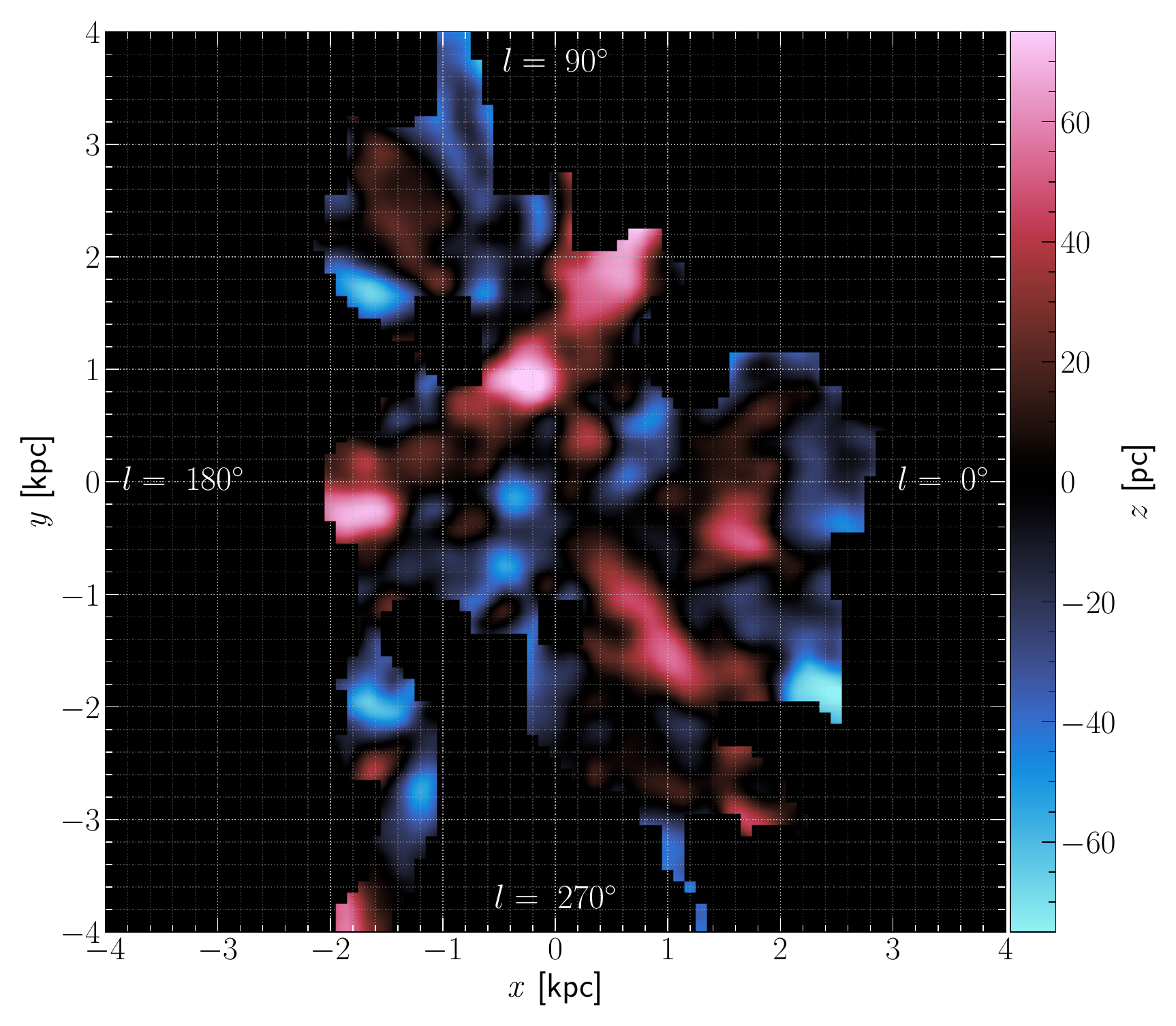}
    \end{minipage}
    \hspace{0.15cm}
    \begin{minipage}{0.485\textwidth}
        \centering
        \includegraphics[width=\textwidth, trim=5 5 0 0, clip]{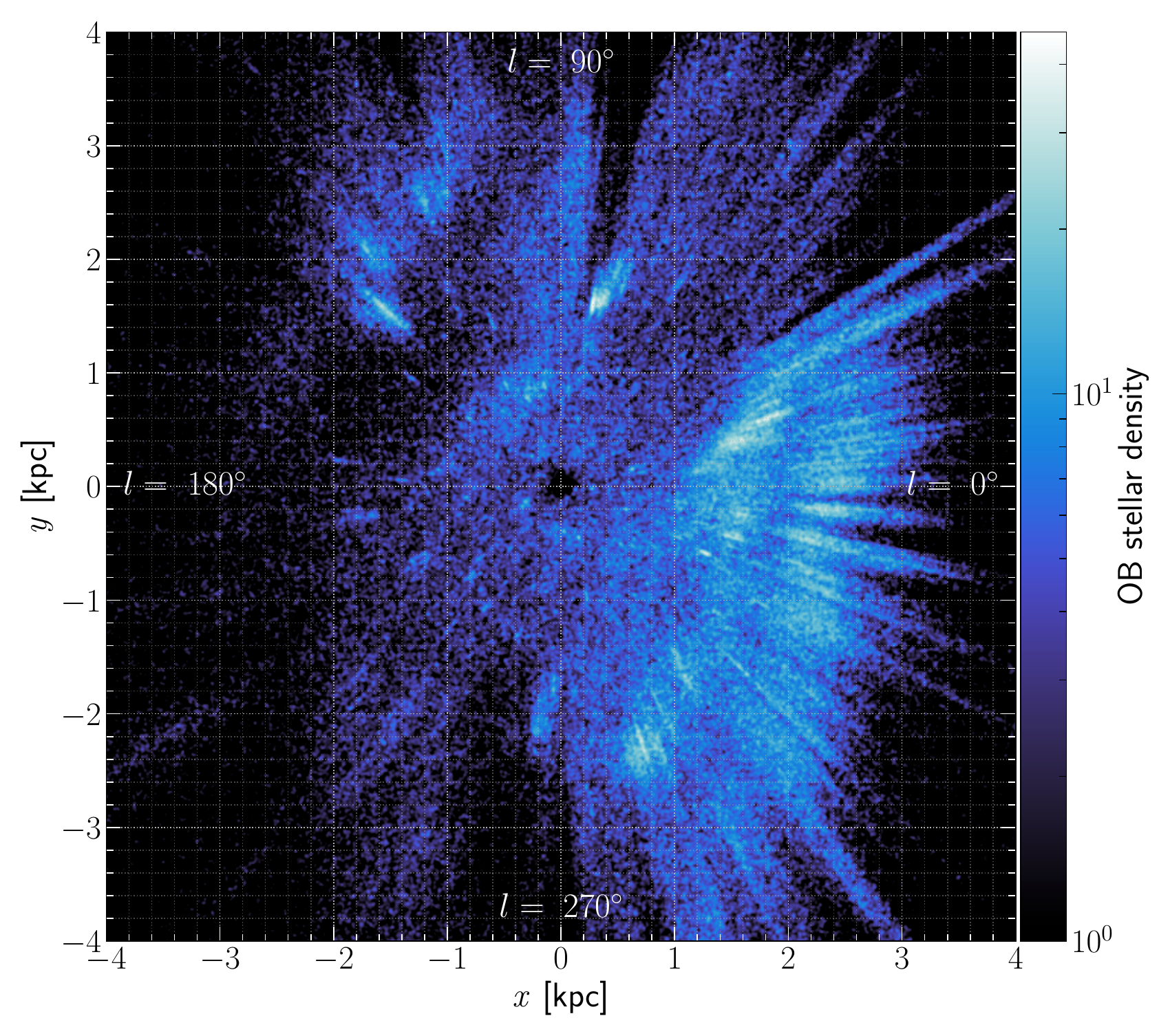}
    \end{minipage}
    
    \caption{Maps of the ${\sim}4$ kpc solar neighbourhood as seen from the north Galactic pole, with the Sun at the center and the Galactic center to the right, using our distance estimates, $d_{\text{OB}}$, for each star. (upper left) Gaussian KDE, with a $25$ pc bandwidth, of the OB stars of the disk ($|z|< 150$ pc) present in the M catalogue. (upper right) Scatter plot of the ALS III stars with good astrometric and photometric quality data in \textit{Gaia} DR3, color-coded according to the photometric categories. Error bars are used to show typical uncertainties for stars located at $1.0$, $2.0$, $2.5$, $3.5$ and $4.5$ kpc, respectively, and can be used to assess how much of the spread in the radial direction is caused by the uncertainties in $d_{\text{OB}}$. (lower left) Gaussian KDE (with a bandwidth of $100$ pc) for the average $z$-height above the mid-Galactic plane for the OB stars of the disk ($|z|< 150$ pc) present in the M catalogue. Positions are vertically shifted by $z_{\odot} = 20$ pc from the plane described by the Galactic equator. Pixels with less than $5$ stars are removed from the map to mask the low-number statistics and avoid an incorrect interpretation of under-dense regions. (lower right) The same as the upper left panel but with distances recalculated for the OBA sample from \citet{Zarietal21}, and a bandwidth of $10$ pc for the KDE.}
    \label{fig:Maps}
\end{figure*}

In this subsection, we present an initial overview of the new Galactic map, shown in the upper left panel of Figure \ref{fig:Maps}, based on our distance estimates for OB stars. A detailed analysis lies beyond the scope of this work, but in the future we will further explore the spatial and kinematical structure of spiral arms, spurs, OB associations and clusters.

Notably, the smaller uncertainties in astrometry provided by \textit{Gaia} DR3 have significantly reduced the radial dispersions of OB associations with respect to the map presented in Paper II. This allows for a more in-focus view of the Perseus and Carina-Sagittarius arms. As stated in Paper II, the Scutum-Centaurus arm seems to be still out of reach, due to the strong extinction towards the inner Galaxy. However, some clusters associated with this arm are easily identifiable. The evidence clearly indicates a physical discontinuity in the density of OB stars in the Perseus arm at Galactic longitudes larger than those of Per~OB1 (\citealt{NeguMarc03} and \citealt{Peeketal22}). This cannot be attributed to interstellar extinction, as the anti-center region is known to have low extinction. The Orion (or Local) arm, where the Sun resides, is the most challenging to interpret, as certain lines of sight in the tangential directions are obscured by larger extinction columns.
The new maps allow for clearer identification and separation of OB associations. In fact, nearly all the associations listed in \citet{Wrig20} can now be distinctly located. Cases like Cygnus OB2, which is now known to include a foreground unrelated association separated by ${\sim}300$ pc (\citealt{Berletal19}), are disentangled when viewed from above on this map.

As an example of the importance of accurate spectral classification for strictly defined OB stars in the context of Galactic cartography, we also present a density map for the OBA sample in \citet{Zarietal21} (lower right panel of Figure \ref{fig:Maps}), using our distance estimates (based on the OB prior for the spatial distribution of young massive stars). Since A-type stars are included in this catalog, and these stars likely dominate in number, the kinematic mixing caused by their long lifetimes washes away small-scale details of the solar neighborhood. This impedes the ability to distinguish structures like the Cepheus and Sagittarius spurs, the separation between the foreground and background associations of Cygnus OB2, and the ring-like structure surrounding the Sun (which we tentatively propose in the following sections may correspond to what has been interpreted as Gould's Belt in the past). However, the OBA sample provides an excellent view of the large-scale structure of the spiral arms, potentially revealing even the Scutum-Centaurus arm. As discussed in section \ref{subsec:needOB}, the sample in \cite{Khaletal24}, composed primarily of late B-type stars, offers a generally cleaner representation than the OBA sample. Figure 9 of that paper presents a map that serves as an intermediate step between the map based on the OBA catalogue and our ALS III map, revealing the Cepheus spur and smaller structures. The catalogue presented in \citet{Quinetal25} is cleaner than the OBA maps from \citet{Zarietal21} and \citet{Chenetal19b}, as evidenced by the less noisy distributions and the comparisons shown in their figure 5. However, by including non-massive ``OB'' stars down to spectral type B9.5, \citet{Quinetal25} still misses important associations, such as Per OB3, that are prominent overdensities in the ALS III map.

Some important features present in this map require particular attention:

\subsubsection{The Cepheus spur}
In Section 3.5 of Paper II, a ``new'' Galactic structure, that we dubbed the Cepheus spur, was described. This inter-arm overdensity, stretching from the Cygnus-Orion arm to the Perseus arm towards the third quadrant, is even more notorious now with the $xy$ maps of the ALS III. However, a historical note is warranted, regarding the primacy of the discovery. Shortly after the publication of Paper II, it was brought to our attention that in the seminal work of \cite{Morgetal53b}, a similar feature had been noted, stating: ``\textit{The aggregates III Cep, I Cam, I Aur, I Gem, and I Mon appear to form a branch to the arm in which the Sun is located}''. They further mentioned the compatibility of this ``\textit{branch}'' with the contemporary radio $21$-cm emission map shown in \citet{Oortetal52}. The structure was suggested sporadically in \cite{Shar65} and again in \cite{Hump70}. After \num{1970}, this suggestion appears to have vanished from the literature, at least as far as the authors are aware, until the publication of Paper II, in 2021. With improved distance measurements for young Galactic clusters, the Galactic maps increased in complexity, making this branching structure less obvious, with some representations of the solar neighbourhood allowing for widely different cartographic interpretations (\citealt{VogtMoff75, Fitz87}). The \textit{Gaia} mission ultimately enabled the mapping of the Cepheus spur using individual stars rather than stellar ``\textit{aggregates}'', inadvertently validating Morgan's old proposition. In Paper II, we demonstrated that the spur is largely kinematically coherent and notable for its distinct anomalous vertical displacement above the mid-Galactic plane, which may yield further insights into the 2D corrugation pattern of the disc. This height anomaly is reinforced with the new map (lower left panel in figure \ref{fig:Maps}).

Since Paper II, several studies have provided further evidence for the existence of the Cepheus spur, using young clusters (\citealt{Alfaetal22, HuntReff23}), 3D dust maps (\citealt{Edenetal24}), and even using A-type stars (\citealt{Ardeetal23}) as tracers. It has also been suggested that the Cepheus spur might be a better match than the Perseus arm for some CO patterns in the longitude-velocity diagrams, and that molecular clouds approximately follow it (\citealt{Peeketal22}). We are currently working on a fourth paper of the Villafranca series of OB groups that will center on the Cepheus spur.

\subsubsection{The Sagittarius spur}
In \citet{Kuhnetal21}, it was shown that the Carina-Sagittarius arm exhibits a branching structure, with a narrow spur extending across the first Galactic quadrant. This was traced using masers, young stellar objects (YSOs), and a 3D dust map. We confirm the existence of the Sagittarius spur as traced by OB stars and extend its path to at least double the originally reported length, continuing through the fourth Galactic quadrant towards Vela OB1. This extension is particularly significant, as it demonstrates the consistency of the structure, as the bias in radial dispersions cannot be used as an argument against the spur's existence where it is traced tangentially to the line of sight.

Interestingly, the Sagittarius spur shares several similarities with the Cepheus spur. Both appear to have a similar pitch angle, remaining largely parallel to each other. Their closest regions to the Sun are coincidentally located at around ${\sim}1$ kpc. Additionally, in the average $z$-height map of the disk (see lower left panel in Figure \ref{fig:Maps}), it is tentatively observed that the Sagittarius spur traces a narrow, nearly straight line, with a slightly higher than expected $z$-height (compared to the surroundings). However, this observation requires a more in-depth analysis for confirmation as the peak in average height is not significant (which is true for the Cepheus spur). If validated, it could support the hypothesis that the Cepheus and Sagittarius spurs are part of a corrugation pattern in the Milky Way disk (\citealt{Teppetal22}).

\subsubsection{Anastasis for Gould's Belt?}

\begin{figure} 
    \centering
    \includegraphics[width=0.48\textwidth, trim = 16 14 0 5, clip]{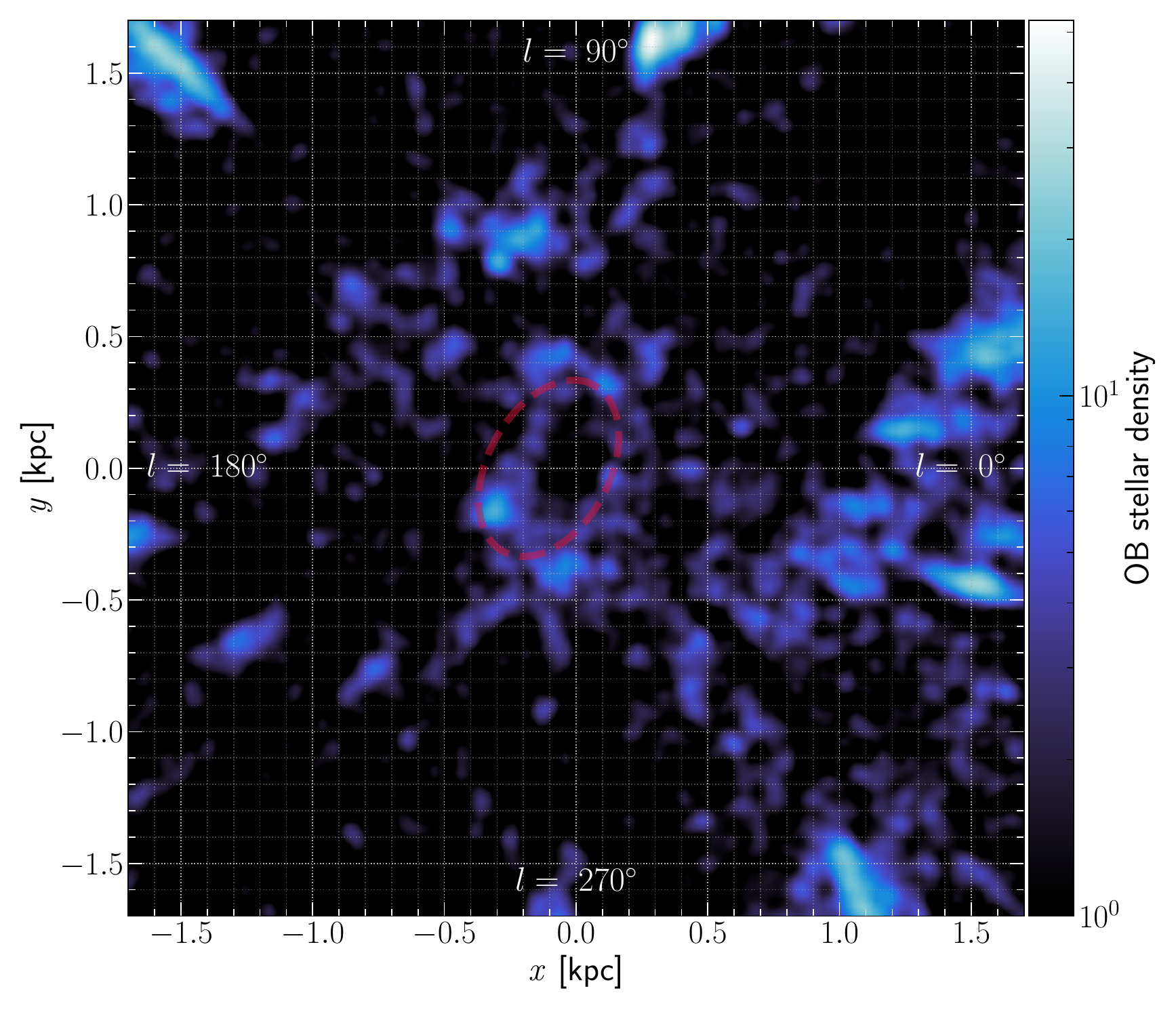}
    \vspace{-4pt}
    \caption{Close-up view of the upper left panel in figure \ref{fig:Maps} showing the $\sim 1.5$ kpc solar neighbourhood. The red dotted ellipse is the best fit model of Gould's Belt from \citet{PerrGren03}.}
    \label{fig:gould}
\end{figure}

For more than 170 years, astronomers have hotly debated about the existence, origin and nature of an evasive structure known as the ``Gould Belt''. First noted by \citet{Hers47} as a band of bright stars visible in the southern skies, it was soon realized that it was close to a great circle around the entire celestial sphere and inclined by some $\sim20^{\circ}$ with respect to the Galactic plane (\citealt{Goul74}). For over a century, various models, such as disks, rings, and even ``sausage''-shaped structures were been proposed with the hope of characterizing this feature with a simple geometrical shape. Observational evidence progressively indicated that OB stars appeared to encircle the Sun within this particular plane, and extending up to about $0.5$ kpc (see \citealt{Popp97} and \citealt{PaloEhle17} for reviews on the topic). The Gould Belt was then said to be expanding (\citealt{Club67}). With the advent of space-based astrometry and the completion of the \textit{Hipparcos} mission, it became possible to use 3D positions and kinematics to investigate the nature of this structure. \citet{Lindetal97} analyzed what was then the most accurate 3D map of OB stars in the solar vicinity. The characteristic inclined plane of Gould's Belt was clearly visible, but no ring-like structure appeared in the top-down view of the Galactic plane. The motions of stars and associations purportedly belonging to the Belt seemed to suggest not only expansion but also rotation, consistent with the differential rotation and tidal stretching expected for a structure of $\sim 1$ kpc in size (\citealt{Olan82}). However, interpreting the Belt's kinematics is challenging, as the solar reflex motion induces a signal in nearby stars that can be easily mistaken for a coherent kinematic pattern. Further analysis strengthened the view that Gould's Belt was a genuine physical structure that could be thoroughly characterized. An age of $\sim 34$ Ma was determined (\citealt{Come99}), and a revised ring model was adopted, specifying its size, inclination, and orientation within the plane of the ellipse (\citealt{PerrGren03}). The history and possible formation of the structure by the high speed collision of an intergalactic cloud (\citealt{Comeetal92}) or even a dark matter clump (\citealt{Bekk09}) were suggested.

Despite accumulating evidence, skepticism about the existence of Gould's Belt persisted. As previously mentioned, the map in \citet{Lindetal97} had failed to show a ring-like structure around the Sun, and the observed kinematics were incoherent enough to be attributed to unrelated star formation episodes (\citealt{Eliaetal09}). An alternative explanation emerged, proposing that the Belt might be a projection effect created by two nearby parallel linear structures, the Radcliffe wave and ``The Split'', without any connecting features bridging the gaps (\citealt{BouyAlve15, Alveetal20}). Hopes that the \textit{Gaia} mission's unparalleled astrometric precision would clarify the structure were dashed, as recent 3D dust maps (\citealt{Edenetal24}) and the OBA star map from \citet{Zarietal21} (lower right panel in Figure \ref{fig:Maps}) show no evidence of the purported Belt. The scientific consensus now leans towards the idea that the Gould Belt was merely a pareidolia in 6D phase space, as all the properties and structural parameters previously attributed to this structure had been defined by considering essentially only two star-forming regions: Orion and Sco–Cen (\citealt{Alfaetal22}).

In Figure \ref{fig:gould} we show that a ring-like structure surrounding the Sun can be clearly seen on the ALS III OB star map when viewed from the north Galactic pole, with stars at large Galactic heights excluded. This ring is largely coincidental with the size and orientation of the purported Gould's Belt, considering the fitted model suggested in \citet{PerrGren03}. The absence of this structure in the OBA map of \citet{Zarietal21} can be readily explained by the dominance of contaminants in that sample (see \ref{subsec:cats}), while the uncertainties in distance measurements from the \textit{Hipparcos} mission were large enough to obscure the Belt's shape in \citet{Lindetal97}. The most prominent associations within this structure are Ori OB1, Vel OB2, Sco-Cen, Lac OB1, and Per OB2/3, with several young stars appearing in between the gaps. It is important to emphasize that \textbf{we are not claiming that the Gould Belt should be reinterpreted as a physically meaningful concept again}; the Radcliffe wave and ``The Split'' (which correspond to the second and fourth quadrant segments of the OB stellar ``Belt'') are moving in opposite directions, partly due to the expansion of the Local Bubble (\citealt{Zucketal22}), and the clusters present in this region are thought to be members of three completely independent families of clusters (\citealt{Swigetal24}), suggesting that the ring is in fact a serendipitous and ephemeral 3D asterism. Additionally, there is an incoherent pattern in the ages (\citealt{Kerretal23}), with the stellar bridges connecting the Radcliffe wave and ``The Split'' (in the first and third Galactic quadrants) likely being older than the aforementioned linear structures (something that could explain the absence of dust in the bridging segments).

Even if the Gould Belt might not be a concept that provides physically meaningful insights, it is still a clearly visible, ring-like pattern of massive stars surrounding the Sun, and not merely a projection effect. We are preparing a follow-up paper on this issue, with a comprehensive spatial and kinematical analysis, aiming to help resolve this debate and point out the weaknesses in the historical arguments that gave rise to the Gould Belt deception.

\section{Future work}

$\,\!$ \indent In this last section we describe the ways in which the ALS project could be extended in the future, by us or by other research groups using it.

\subsection{Existing \textit{Gaia} data and ongoing surveys}

$\,\!$ \indent As we have already mentioned, one of our immediate goals is to extend the ALS catalogue to the LMC, SMC, and MB. From the point of view of the selection of sources using \textit{Gaia} this is relatively easy for most of the sources using a combination of parallaxes and proper motions \citep{Lurietal21}, especially as most of the stars in the Magellanic Clouds have low extinctions. However, \textit{Gaia} should be complemented with high-resolution studies in crowded fields such as R136 \citep{Crowetal16,Bestetal20}. We have already started to compile the literature data in a similar way to what was done for the ALS decades ago and we will include spectroscopy from GOSSS, LiLiMaRlin, VFTS, XShootU, and other future surveys. \textit{Gaia} will also be used to provide information on variability \citep{Maizetal23}.

A sometimes ignored characteristic of the \textit{Gaia} photometric system is that the passbands for a given filter change slightly from one data release to another (see \cite{Maiz17a,MaizWeil18} and compare it with the nominal one in \cite{Jordetal10}). The cause of this is double: there was frozen water in some optical elements at the beginning of the mission and the photometry is processed independently for each data release, with some electronic effects created by the use of TDI gates that generate magnitude-dependent effects with jumps \citep{Prusetal16}. As we discuss in \citet{MaizWeil24}, there is a significant difference in the sensitivity of \GBP\ between DR2 and EDR3 to the left of the Balmer jump for stars with $\GG > 10.87$~mag. In a future publication (Weiler et al. in prep.) we will present a full recalibration of the \textit{Gaia} 6-filter system ($G_{\rm BP2}+G_{\rm 2}+G_{\rm RP2}+G_{\rm BP3}+G_{\rm 3}+G_{\rm RP3}$) that will yield an unprecedented accuracy for a whole-sky optical photometric survey. The unexpected effect regarding the Balmer jump can be exploited to our advantage to provide \Teff\ estimates for low-extinction OB stars by measuring $G_{\rm BP3} - G_{\rm BP2}$ (Fig.~7 in \citealt{MaizWeil24}) and using colour-colour plots analogous to the classical $U-B$ vs. $B-V$ ones of \citet{JohnMorg53}. Therefore, one of the lines we will explore will involve the use of this technique to discriminate the nature of the ALS sample and expand it in a systematic way.

Another line of work to be explored is the combination of \textit{Gaia} and 2MASS photometry to create colour-colour and colour-absolute magnitude diagrams that can be used to detect hidden OB stars, a technique exploited by \citet{Zarietal21}. The difficulty arises from the fact that for $\GBP-\GRP > 1.0$~mag the population of extincted OB stars is a small minority in a sea of red giants and supergiants. The new \textit{Gaia} calibration described in the previous paragraph will be used in this respect to yield a cleaner sample of extincted OB stars that are not currently present in the ALS.

A different approach is to use the XP low-resolution spectra combined with other information from \textit{Gaia} to determine \Teff\ and extinction for a large sample and, from there, extract the OB stars \citep{Creeetal23a}. However, the fact that the extinction law is not uniform between sightlines severely complicates the problem and can lead to erroneous results in the determination of both \Teff\ and extinction \citep{Maiz24}. Furthermore, the peculiar characteristics of the \textit{Gaia} low-resolution spectrometers introduce complications in its use for synthetic photometry extracted from it \citep{Weiletal20} and the measurement of individual spectral lines require specific techniques \citep{Weiletal23}. Indeed, \textit{Gaia} XP spectrophotometry is not provided as one would in principle expect, a table of flux as a function of wavelength, but as coefficients for a series of basis functions in wavelength \citep{DeAnetal23,Montetal23a}, which can introduce artifacts in the result if not done correctly. For those reasons, XP spectrophotometry holds great promise but one must be careful when analyzing its results.

The three more immediate science goals that will be tackled with the current version of the ALS will be an analysis of the Gould's belt, a study of the Galactic rotation curve of OB stars (and the possible non-axisymmetric residuals) and a search for massive runaways and walkaways. Regarding the rotation curve, M.P.G. recently wrote his Masters thesis on the issue and a paper is forthcoming. For runaways and walkaways we did a preliminary analysis using \textit{Gaia}~DR1 \citep{Maizetal18b} that was followed by searches around specific clusters \citep{Maizetal22b, Maizetal24a}. Other groups have used GOSC to do similar analyses \citep{Carretal23}. With the new improved sample and the specific Galactic rotation curve we expect to find many more.\\

Regarding ongoing surveys, as we already mentioned in the introduction, we are using our GOSSS and LiLiMaRlin surveys to obtain spectra of the ALS sources in order to confirm their nature and determine their properties. Here we detail our ongoing efforts in this direction and in others using surveys we are leading:

\begin{itemize}
 \item Since 2014 we have an ongoing filler program at the Gran Telescopio de Canarias (GTC) to obtain intermediate-resolution spectroscopy of faint OB stars using the OSIRIS instrument. This was initially under the umbrella of GOSSS but it later became an ALS-specific program. The big advantage of this program is the use of a 10.4~m telescope.
 \item The LiLiMaRlin project includes both public and private high-resolution spectroscopy. Among the private data it includes two of our prior observing programs, CAF\'E-BEANS using the 2.2~m Telescope at Calar Alto \citep{Neguetal15a} and NoMaDS using the 10~m Hobby-Eberly Telescope. It also incorporates the spectra of the IACOB project (\citealt{SimDetal11a}; see also \citealt{Holgetal20, Holgetal22} and \citealt{deBuetal23, deBuetal24} for recent updates regarding the O-type and B supergiant samples, respectively) obtained with the 2.56~m Nordic Optical Telescope and 1.2~m The Mercator Telescope at La Palma. In the last years we have been granted time with those two telescopes and with the 3.6~m Telescopio Nazionale Galileo (also at La Palma) for specific ALS programs with the objective of observing all ALS stars brighter than $B = 10.5$~mag that can be reached from La Palma. Our goal is to observe them by the end of 2027. 
 \item We have also started a program with the WEAVE LIFU\footnote{As of the time of this writing the WEAVE MOS mode is not yet available but is expected to be soon.} to obtain medium-resolution spectroscopy of compact northern clusters with OB stars. A LIFU is the optimal solution in such circumstances, as it allows the obtention of simultaneous spectroscopy of several sources where single- or multi-fibre instruments would allow only for one and long-slit instruments of two or maybe three. 
 \item In the southern hemisphere we have gathered a large sample of public high-resolution spectra obtained with FEROS at the 2.2~m La Silla MPG Telescope, HARPS at the 3.6~m La Silla Telescope, UVES at the 8.2~m VLT/UT2, and with the HRS at the 9.2~m South African telescope. At the same time, we are conducting a survey of ALS stars using FEROS.
 \item We also plan to exploit our GALANTE survey \citep{LorGetal19,LorGetal20,Maizetal21d}, which conducted a seven narrow- and intermediate-band filter photometric survey of a fraction of the northern Galactic plane, concentrating in areas with large numbers of OB stars. GALANTE complements \textit{Gaia} photometry, especially in crowded and \HII\ regions, and is being followed up by MUDEHaR \citep{Holgetal24} to analyze the variability of the O stars in the continuum and in H$\alpha$. 
 \item All of the above data are being used to study three complementary aspects of massive stars whose results will be eventually incorporated into the ALS web site. \textbf{1. Multiplicity} is being analyzed in the northern hemisphere through the MONOS project \citep{Maizetal19b,Trigetal21}, to be followed by its southern counterpart MOSOS soon. \textbf{2. Group membership} in clusters and (sub)associations is being carried out through the Villafranca project \citep{Maizetal20b,Maizetal22a,Maizetal24b}. \textbf{3. The ISM in their sightlines} is being analyzed with the CollDIBs project and through the study of the extinction laws \citep{Maiz15a,MaizBarb18,Maizetal21a}. 
\end{itemize}

\subsection{Next \textit{Gaia} data releases and future surveys}

$\,\!$ \indent As of the time of this writing\footnote{See \url{https://www.cosmos.esa.int/web/gaia/release}.}, two additional \textit{Gaia} data releases are planned: DR4, with 66 months of data ``not before mid 2026'', and DR5, with all mission data, ``not before the end of 2030''. What improvements for the ALS should we expect based on them? The most straightforward one would be a reduction of random and systematic uncertainties for all types of data: astrometric, photometric, and spectroscopic. That should lead to a better characterization of the Galactic structure and kinematics defined by OB stars.

Besides the reduction in the uncertainties, new data should become available. Currently there is only low-resolution spectrophotometry and Radial Velocity Spectrometer data for a fraction of the sample and with the new data releases those should become available for a majority of the stars above a certain magnitude threshold. Similarly, epoch photometry should become available for a much larger and less biased sample, leading to the generation of light curves for the study of variable OB stars such as eclipsing binaries and OBe stars. Finally, the possibly most interesting results will originate in the epoch astrometry and radial velocity measurements, which, in combination with ground-based spectroscopy, should lead to a thorough analysis of binary systems (especially those with compact objects). 

$\,\!$ \indent Looking down the road, the next decade will see the start and exploitation of two large multi-fibre spectroscopic surveys, WEAVE in the north and 4MOST in the south, that will produce high-quality data for tens of thousands of OB stars and whose results will be incorporated into the ALS. Two decades from now, the revolution should be GaiaNIR\footnote{\url{https://www.astro.lu.se/GaiaNIR}.}, which will hopefully open the door to the population of heavily extincted OB stars in the distant Galactic plane and significantly improve the accuracy of proper motions by combining data from the two missions.

\section*{Acknowledgements}

$\,\!$\indent We dedicate this paper to Rodolfo Barb\'a, our dearest friend and colleague, who passed away after significantly contributing to it but before its completion. Even though he is not on this Earth anymore, we think he would have enjoyed the final result. We thank Robert Benjamin for his comments on the Cepheus spur and the history of Milky Way cartography. M.P.G. and J.M.A. acknowledge support from the Spanish Government Ministerio de Ciencia e Innovaci\'on and Agencia Estatal de Investigaci\'on (\num{10.13039}/\num{501100011033}) through grants PGC2018-0\num{95049}-B-C22 and PID2022-\num{136640}~NB-C22 and from the Consejo Superior de Investigaciones Cient\'ificas (CSIC) through grant 2022-AEP~005. This work has made use of data from the European Space Agency (ESA) mission {\it Gaia} (\url{https://www.cosmos.esa.int/gaia}), processed by the {\it Gaia} Data Processing and Analysis Consortium (DPAC, \url{https://www.cosmos.esa.int/web/gaia/dpac/consortium}).  Funding for the DPAC has been provided by national institutions, in particular the institutions participating in the {\it Gaia} Multilateral Agreement. The {\it Gaia} data are processed with the computer resources at Mare Nostrum and the technical support provided by BSC-CNS. This research has made extensive use of the \href{http://simbad.u-strasbg.fr/simbad/}{SIMBAD} and \href{https://vizier.u-strasbg.fr/viz-bin/VizieR}{VizieR} databases, operated at \href{https://cds.u-strasbg.fr}{CDS}, Strasbourg, France. 

\section*{Data availability} 

$\,\!$\indent The data from this paper are available from the CDS and from our web site, \url{https://als.cab.inta-csic.es}, which will soon supersed the GOSC database, \url{https://gosc.cab.inta-csic.es}. The astrometric and photometric calibrations, along with the distance estimation algorithm, are publicly available for community use in Python through the following GitHub repository: \url{https://github.com/MichelangeloPantaleoni/Gaia-DR3-Corrections-and-recipes}.



\bibliographystyle{mnras}
\bibliography{general} % if your bibtex file is called example.bib

% Don't change these lines
\bsp	% typesetting comment
\label{lastpage}
\end{document}